\documentclass[aps,pra,reprint,superscriptaddress]{revtex4-2}

\usepackage{graphicx}      
\usepackage{dcolumn}       
\usepackage{bm}            
\usepackage{xcolor}        
\usepackage{amsmath, amssymb} 
\usepackage{mathrsfs}      
\usepackage[percent]{overpic} 
\usepackage{adjustbox}     
\usepackage{pifont}        
\usepackage{float}         
\usepackage{hyperref}
\hypersetup{    
colorlinks=true,
    linkcolor=blue,
    citecolor=blue,      
    urlcolor=blue
}
\usepackage{placeins}

\usepackage{silence}
\newif\ifshownotes
\shownotesfalse   

\renewcommand{\lg}[1]{
  \ifshownotes
    {\color{orange}\bf LG: {#1}}
  \fi
}

\newcommand{\cmark}{\ding{51}}  
\newcommand{\xmark}{\ding{55}}

\newcommand{\revA}[1]{#1}   
\newcommand{\revB}[1]{#1}   
\begin{document}
	
\title{Interference between multiple photoionization pathways in chiral molecules: Converging continuum results in Gaussian bases}
	
\author{Muhammad Sakhi}
\affiliation{J. R. Macdonald Laboratory, Department of Physics, Kansas State University, Manhattan, Kansas 66506, USA}

\author{Alexander Blech}
\affiliation{Freie Universit\"at Berlin, Fachbereich Physik \& Dahlem Center for Complex Quantum Systems, Berlin, Germany}

\author{Corbin Allison}
\affiliation{J. R. Macdonald Laboratory, Department of Physics, Kansas State University, Manhattan, Kansas 66506, USA}

\author{Christiane P. Koch}
\affiliation{Freie Universit\"at Berlin, Fachbereich Physik \& Dahlem Center for Complex Quantum Systems, Berlin, Germany}

\author{Loren Greenman}
\email{lgreenman@ksu.edu}
\affiliation{J. R. Macdonald Laboratory, Department of Physics, Kansas State University, Manhattan, Kansas 66506, USA}

\date{\today}
	
\begin{abstract}
An accurate description of photoionization observables is a central challenge
for theoretical models of molecular photoionization.  Within standard
electronic-structure approaches, the continuum states are represented by unoccupied
Hartree-Fock orbitals expanded in Gaussian basis sets. Since these
basis sets are optimized for bound states, computed observables may exhibit a
noticeable basis-set dependence.  Here, we augment these basis sets with
diffuse functions and investigate the convergence of photoelectron circular
dichroism (PECD), anisotropy parameters, and the forward--backward ionization
time delay in the multiphoton ionization of randomly oriented \revA{methyloxirane
molecules}. All observables converge systematically with the number of added
diffuse functions, resolving previously observed basis-set discrepancies and
indicating that the augmented basis sets provide a more accurate representation
of the intermediate continuum states. In particular, our fully \textit{ab
initio} calculations yield forward--backward time delays in qualitative
agreement with recent measurements, for which previous theoretical descriptions
relied on empirical modeling. Diffuse augmentation thus provides a
computationally efficient and transferable route to converged Gaussian-basis
calculations for molecular multiphoton photoionization.
\end{abstract}
	
\maketitle
	
\section{\label{sec1}Introduction}
Photoionization is one of the most fundamental light--matter interaction
processes and a primary probe of electronic structure and dynamics in atoms
and molecules~\cite{pazourek2015RMP, nisoli2017ChemRev}. Advances in the
generation of attosecond and extreme ultraviolet (XUV) light pulses have
turned it into a time-resolved probe of electron dynamics, enabling the
measurement of photoemission time delays~\cite{schultze2010Science, klunder2011PRL, isinger2017Science, han2025Nature} and inspiring theoretical
proposals for controlling the electronic coherence of the
photoion~\cite{goetz2016PRA}.

Among the molecular properties that photoionization gives access to is
chirality---the inability of a molecule to be superimposed on its mirror
image (enantiomers)~\cite{mcnaught1997GoldBook}. Chirality plays an
important role in chemistry, biology, and pharmacology~\cite{ma2020CSBB,
  liu2015ChemRev}.
While many spectroscopic methods do not distinguish between enantiomers, techniques using circularly polarized light break inversion symmetry and can distinguish chiral molecules.

A prominent example is PECD, the
difference in the angular distribution of photoelectrons emitted from
randomly oriented chiral molecules upon ionization with left- and
right-circularly polarized light~\cite{ritchie1976PhysRevA}. It arises from electric-dipole interactions and produces a strong asymmetric signal that is highly sensitive to molecular chirality~\cite{nahon2015JESRP, fiechter2023StructDyn, powis2000JCP, boewering2001PRL,waters2022CPC, lux2012ACIE, lux2015CPC, lux2016JPhysB, kastner2016CPC,
  kastner2017JCP, lehmann2013JCP, janssen2014PCCP, fanood2015NatCom}. PECD has been investigated by employing ionization techniques on chiral
molecules~\cite{lux2012ACIE, lux2015CPC, lux2016JPhysB, hergenhahn2004JCP,
goetz2017JCP, beaulieu2017Science}, and perturbation theory has been used to
explain it~\cite{goetz2019JCP, goetz2019PRL,
  goetz2025PRR, han2025Nature}. PECD probes molecular chirality solely through interactions with electric fields, making it an effective platform for coherent control schemes such as pump-probe spectroscopy~\cite{tikhonov2022SciAdv}, pulse shaping~\cite{goetz2019PRL}, and carrier envelope phase control~\cite{hanus2023RSC}. Recently, a photoelectron interferometric scheme based on the RABBITT (reconstruction
of attosecond beating by interference of two-photon transitions) technique
enabled coherent control of PECD and the separation of bound- and continuum-state contributions in randomly oriented chiral molecules~\cite{goetz2025PRR}. Advances in the generation of circularly polarized extreme XUV radiation made its experimental realization possible~\cite{han2025Nature}. In this scheme, the chiral observables are built from interfering two-photon ionization pathways and thus involve transitions between continuum states.

The theoretical description of multiphoton ionization requires an accurate treatment of both the localized bound electronic states of the molecule and the delocalized continuum states of the emitted photoelectron. Time-dependent perturbation theory provides a rigorous framework for describing weak-field multiphoton ionization processes relevant to attosecond interferometric techniques. Its predictive capability, however, depends critically on an accurate representation of the electronic continuum~\cite{douguet2018PRA}. Dedicated continuum approaches, including the complex Kohn variational method~\cite{douguet2018PRA} and B-spline or single-center expansion methods~\cite{stener2004JCP, demekhin2015JCP}, satisfy this requirement but become computationally demanding for complex molecules. In contrast,
approaches built on standard quantum-chemistry Gaussian basis sets~\cite{goetz2017JCP, goetz2019PRL} are simple and applicable to molecules of arbitrary structure. 
However, Gaussian basis sets are optimized to describe bound states, and it must be verified that they can also represent continuum-like states with sufficient accuracy.

A particular difficulty arises when transition dipole matrix elements between continuum orbitals need to be calculated, as in the case of the RABBITT two-photon process~\cite{goetz2025PRR,han2025Nature}. These are hard to evaluate because the continuum orbitals are spatially delocalized.
As a remedy the intermediate continuum states of the RABBITT process can be represented as a summation over unoccupied Hartree--Fock orbitals~\cite{goetz2025PRR,han2025Nature}, exploiting the fact that the dipole-perturbed states remain semi-localized and can thus be expanded in a localized Gaussian basis. This representation holds at low perturbation orders, since each action of the dipole operator extends the perturbed state incrementally, and its success depends on the convergence of the Gaussian basis used to construct the Hartree--Fock orbitals.
While the single-photon PECD agrees well between the aug-cc-pVTZ and aug-cc-pVQZ basis sets~\cite{goetz2025PRR}, the two-photon PECD shows noticeable discrepancies between them. These deviations likely arise from the limitations of standard quantum chemistry Gaussian basis sets. These basis sets are well suited for representing  localized bound states, where the electronic wave-function is concentrated near the nuclei, but not the continuum states. Moreover, the unoccupied Hartree–Fock orbitals obtained with these basis sets may not be energetically dense enough to accurately describe the intermediate continuum states within the time-dependent perturbation theory framework.
\raggedbottom

Here, we study the multiphoton ionization of chiral molecules within time-dependent perturbation theory and establish how standard Gaussian basis sets must be augmented with diffuse functions to obtain a converged description of the underlying continuum--continuum transitions. We consider time-resolved RABBITT-PECD within the framework of Refs.~\cite{goetz2019PRL, goetz2025PRR, Blech2025} for the randomly oriented prototypical chiral molecule methyloxirane (C$_3$H$_6$O), which was also the focus of the experimental work in Ref.~\cite{han2025Nature}. \revA{Its electronic excited states were recently characterized by high-resolution vacuum ultraviolet (VUV) photoabsorption measurements combined with time-dependent density functional
theory (TDDFT) calculations in 6.0--10.8\,eV photon energy range~\cite{MendesPCCP2026}. In this work,} we systematically monitor the convergence of the multiphoton PECD, the photoelectron anisotropy parameters, and the forward--backward photoionization time delay with respect to the number of added diffuse functions, showing that the added functions substantially improve the representation of the intermediate continuum states and resolve the basis-set discrepancies observed in Ref.~\cite{goetz2025PRR}. The forward--backward photoionization time delay constitutes the most rigorous of these tests: as an enantiosensitive, attosecond-scale observable, it is sensitive to the phases of the two-photon transition amplitudes, so that basis set errors in the continuum representation appear directly as spurious delays. Its systematic convergence thus validates the approach at the level of spectral phases, and our converged delays qualitatively reproduce the measurements of Ref.~\cite{han2025Nature}. Our results thus bring the time-resolved modeling of attosecond photoionization dynamics in polyatomic molecules within reach of standard quantum-chemistry machinery.

\section{\label{sec2}Theoretical Approach}
We consider photoionization of a molecule by an attosecond pulse train in the presence of a phase-locked dressing field, as realized in the RABBITT technique~\cite{paul2001Science, muller2002APB}. 
\revB{The electric field is
taken as the superposition of an XUV frequency comb and the IR field,
\begin{equation}
	\bm{\mathcal{E}}(t) = \bm{\mathcal{E}}^{(\mathrm{XUV})}(t)
	+ \bm{\mathcal{E}}^{(\mathrm{IR})}\!\left(t - \tau_{\mathrm{IR}}\right),
	\label{field1}
\end{equation}
with
\begin{equation}
	\bm{\mathcal{E}}^{(\mathrm{XUV})}(t) = \sum_{j=1}^{N_p} \mathcal{E}_j\,
	e^{-\frac{(t-\tau_j)^2}{2\sigma_j^2}}\,
	\mathrm{Re}\!\left\{ e^{-i\left[\omega_j (t - \tau_j) + \phi_j\right]}\,
	\mathbf{e}_{\mathrm{XUV}} \right\},
	\label{field2}
\end{equation}
where $j = 2q + 1$ with $q$ an integer, $N_p$ is the number of harmonic components in the XUV comb, and
$\mathcal{E}_j$, $\omega_j$, $\phi_j$, and $\tau_j$ are the amplitude,
frequency, phase, and time delay of the $j$th component, respectively. The
Gaussian envelope of each component has an intensity full width at half maximum (FWHM) of
$2\sqrt{\ln 2}\,\sigma_j$, and
$\mathbf{e}_{\mathrm{XUV}} = (\hat{\mathbf{x}} \pm i\hat{\mathbf{y}})/\sqrt{2}$
is the circular polarization unit vector of the XUV field. The comb consists of odd harmonics of a fundamental frequency $\omega_0$, with photon energy
$\hbar\omega_{2q+1} = (2q+1)\hbar\omega_0$.}
One-photon absorption from the harmonic $\mathcal{H}_{2q+1}$ produces photoelectrons with kinetic energy
$\epsilon_{k} = \hbar\omega_{2q+1} - |\epsilon_{0}|$, where
$|\epsilon_{0}|$ denotes the ionization potential, giving rise to the harmonic peaks of the photoelectron spectrum. An infrared (IR) field with mean photon energy $\hbar\omega_{\mathrm{IR}} = \hbar\omega_{0}$ opens two-photon pathways: IR absorption following ionization by $\mathcal{H}_{2q-1}$ and IR emission following ionization by $\mathcal{H}_{2q+1}$. Their interference at the same final energy gives rise to sidebands $\mathcal{S}_{2q}$ between consecutive harmonic peaks.

To describe these dynamics, we follow the framework introduced in
Refs.~\cite{goetz2019JCP, goetz2019PRL, goetz2025PRR, Blech2025} and treat the photoionization within the fixed-nuclei, nonrelativistic dipole approximation. The many-body wave function is written as a linear combination of the Hartree--Fock ground state, singly excited bound states, and continuum states,
\begin{align}
	|\Psi^N(t)\rangle &= \alpha_{0}(t)e^{-i\epsilon_{0}t}|\Phi_{0}\rangle
	+ \sum_{i,a} \alpha_{i}^{a}(t)e^{-i\epsilon_{i}^{a} t}|\Phi_i^a\rangle \nonumber \\
	&\quad + \sum_i \int dk \, \alpha_{i}^{k}(t)e^{-i\epsilon_{i}^{k}t} |\Phi_i^k \rangle ,
	\label{wavefunction}
\end{align}
where $|\Phi_i^a\rangle$ ($|\Phi_i^k\rangle$) denotes a particle--hole excitation from the occupied orbital $i$ into the unoccupied orbital $a$ (continuum orbital $k$), and $\alpha_{0}(t)$, $\alpha_{i}^{a}(t)$, and $\alpha_{i}^{k}(t)$ are the corresponding time-dependent coefficients. Following Ref.~\cite{goetz2019PRL}, we obtain $|\Psi^{N}(t)\rangle$ by
solving the time-dependent Schr\"odinger equation describing the
interaction of the molecule with the combined XUV and IR fields.

The PECD is determined by the difference in the photoelectron angular distribution (PAD),
\begin{equation}
	\mathrm{PECD}(\epsilon_{k},\Omega_k) =  	\frac{1}{\mathcal{N}}\Bigg(\frac{d^{2}\sigma^{(\pm)}}{d\epsilon_k\, d\Omega_k} - 	\frac{d^{2}\sigma^{(\mp)}}{d\epsilon_k\, d\Omega_k}\Bigg).
	\label{pecdeq}
\end{equation}
Here, $\pm$ denotes the field polarization, $k$ the photoelectron momentum, $\epsilon_{k}$ the photoelectron energy, and $\Omega_k=(\theta_k,\phi_k)$ the solid angle of the photoelectron momentum vector. $\mathcal{N}$ is a normalization factor. We normalize the PECD with respect to the maximum intensity of the PAD. We obtain the photoelectron angular distribution by averaging over all molecular orientations,
\begin{equation}
	\frac{d^{2}\sigma}{d\epsilon_k\, d\Omega_k} \approx \int \left| \alpha_{i_0}^{(1)k}(t;\gamma_{\mathscr{R}}) + \alpha_{i_0}^{(2)k}(t;\gamma_{\mathscr{R}}) \right|^2 \, d^3\gamma_{\mathscr{R}}~,
	\label{oiret_avg_PAD}
\end{equation}
where $\alpha_{i_{0}}^{(1,2)k}(t;\gamma_{\mathscr{R}})=\langle \Phi_{i_0}^{k}|\Psi_{\gamma_{\mathscr{R}}}^{N(1,2)}(t) \rangle$ are the first- (second-) order correction in the limit $t \rightarrow \infty$.  Here, $i_{0}$ denotes the initially occupied
Hartree--Fock orbital from which ionization occurs, \revA{which in this study is
the highest occupied molecular orbital (HOMO);} $k$ labels the photoelectron momentum of the continuum orbital, and $\gamma_{\mathscr{R}}=(\alpha,\beta,\gamma)$ are the Euler angles with respect to the laboratory frame. The state $|\Phi_{i_{0}}^{k}\rangle$ describes a particle–hole excitation into the continuum orbital $\phi^{-}_{k}(r_N)$ and is constructed as antisymmetrized product of the $(N-1)$-electron ionic component and the photoelectron scattering wave function $\phi^{-}_{k}(r_N)$~\cite{McCurdy2001PRA63, McCurdy2001PRA64}, which we obtain by solving the scattering problem~\cite{Lucchese1982PRA}. \revB{The angle and energy resolved PAD can be written in terms of the contributions from one- and two-photon ionization processes,
\begin{equation}
	\frac{d^{2}\sigma^{(\pm)n\mathrm{ph}}}{d\epsilon_k\, d\Omega_k} = \int  \alpha_{i_0}^{(n)k}(t;\gamma_{\mathscr{R}}) \alpha_{i_0}^{*(n)k}(t;\gamma_{\mathscr{R}})  \, d^3\gamma_{\mathscr{R}}~,
	\label{PAD1ph2ph}
\end{equation}
where $n = 1,2$. We further expand the PAD in terms of associated Legendre polynomials $P_{L}^{M}$,

\begin{equation}
\frac{d^{2}\sigma^{(\pm)n\mathrm{ph}}}{d\epsilon_k\, d\Omega_k}= \sum_{L,M} \beta_{LM}^{(\pm)(n\mathrm{ph})}(\epsilon_{k})P_{L}^{M}(\cos\theta)e^{iM\phi} ,
\label{padeq}
\end{equation}}
where $\beta_{LM}(\epsilon_{k})$ are the anisotropy parameters. Since the PAD is a real-valued observable, the complex anisotropy parameters must satisfy the symmetry relation
\begin{equation}
 \beta_{LM}(\epsilon_{k})^{*}=(-1)^{M}\beta_{L-M}(\epsilon_{k}).
 \label{betasym}
\end{equation}
Each $\beta_{LM}$ receives contributions from one-photon and two-photon processes:
\begin{equation}
	\beta_{LM}(\epsilon_{k}) =
	\beta_{LM}^{(1\mathrm{ph})}(\epsilon_{k})
	+ \beta_{LM}^{(2\mathrm{ph})}(\epsilon_{k}) .
	\label{betasum}
\end{equation}
Here, $\beta^{(1\mathrm{ph})}_{10}$ is determined from the first-order
perturbative correction $\alpha^{k(1)}_{i_0}(t)$, which describes
single-photon ionization from the initial bound state $|\phi_{i_0}\rangle$
into the continuum state $|\phi_{k}^{-}\rangle$,
\begin{equation}
	\alpha^{k(1)}_{i_0}(t)
	= i\langle \phi_{k}^{-} | \hat{r} | \phi_{i_0} \rangle
	\int_{-\infty}^{t} dt'\,
	e^{i(\varepsilon_{k}-\varepsilon_{i_0}) t'}
	E(t').
	\label{1stordercorr}
\end{equation}
\revB{Transforming Eq.~\eqref{1stordercorr} into the laboratory frame, substituting
it into Eq.~\eqref{PAD1ph2ph}, and comparing the result with
Eq.~\eqref{padeq}, we obtain the orientation-averaged laboratory-frame
one-photon anisotropy parameters,
\begin{equation}
	\begin{split}
		&\beta_{L,M}^{(1\mathrm{ph})}(k)
		= (-1)^{M}\sum_{\mu_{0},\mu_{0}'}(-1)^{\mu_{0}}\,
		F_{\mu_{0}}^{(1)\,i_{0}}(k,\infty)\,
		F_{\mu_{0}'}^{(1)\,i_{0}}(k,\infty)^{*}\\
		&\times\sum_{l,m,\mu}\sum_{l',m',\mu'}(-1)^{\mu'+m'}\,\eta(l,l',L)
		I_{l,m,\mu}^{(1)\,i_{0}}(k)\, \\
		&\times I_{l',m',\mu'}^{(1)\,i_{0}}(k)^{*}
		\begin{pmatrix} l & l' & L \\ 0 & 0 & 0 \end{pmatrix}
		\begin{pmatrix} l & l' & L \\ m & -m' & m'-m \end{pmatrix}\\
		&\times\begin{pmatrix} 1 & 1 & L \\ \mu & -\mu' & m'-m \end{pmatrix}
		\begin{pmatrix} 1 & 1 & L \\ -\mu_{0} & \mu_{0}' & -M \end{pmatrix},
	\end{split}
	\raisetag{\baselineskip}
	\label{eq:beta1ph}
\end{equation}
with
\begin{equation}
	\begin{aligned}
		F_{\mu_0}^{(1)\,i_0}(k,t) &\equiv \int_{-\infty}^{t} dt'\,
		E_{\mu_0}(t')\, e^{i(\epsilon_k - \epsilon_{i_0})t'}, \\
		I_{l,m,\mu}^{(1)\,i_0}(k) &\equiv
		\langle \varphi_{k,l,m}^{-} | \hat{d}'_{\mu} | \varphi_{i_0} \rangle
		= I_{k,l,m,\mu}^{i_0}, \\
		\eta(l,l',L) &\equiv \sqrt{\frac{(2l+1)(2l'+1)(2L+1)}{4\pi}}.
	\end{aligned}
	\label{eq:defs1ph}
\end{equation} }
Similarly, $\beta^{(2\mathrm{ph})}_{LM}$ is determined from the second-order perturbative correction $\alpha^{k(2)}_{i_0}(t)$, which describes XUV+IR two-photon ionization to the final continuum state.
\begin{equation}
	\begin{split}
		&\alpha^{k(2)}_{i_0}(t)
		= -\langle \phi_{k}^{-} | \hat{r} | \phi_{i_0} \rangle
		\sum_{i\in \mathrm{occ}} \langle \phi_i | \hat{r} | \phi_i \rangle
		\int_{-\infty}^{t} dt'\, e^{i(\varepsilon_{k} - \varepsilon_{i_0})t'} E(t')\\
		&\times\int_{-\infty}^{t'} dt''\, E(t'') 
		- \int dk_{1} \sum_{i\in \mathrm{occ}}
		\langle \phi_{k}^{-} | \hat{r} | \phi_{k_1} \rangle
		\langle \phi_{k_1} | \hat{r} | \phi_{i} \rangle\\
		&\times\int_{-\infty}^{t} dt'\, e^{i(\varepsilon_{k} - \varepsilon_{k_1})t'} E(t') 
		 \int_{-\infty}^{t'} dt''\,
		e^{i(\varepsilon_{k_1} - \varepsilon_{i})t''} E(t''),
	\end{split}
	\raisetag{3\baselineskip}
	\label{2ndordercorr}
\end{equation}
where $|\phi_{i}\rangle$ is the initial bound state, $|\phi_{k_1}\rangle$ are the intermediate continuum states (accessed after XUV-photon absorption), and $|\phi_{k}^{-}\rangle$ is the final continuum state of the photoelectron in the RABBITT scheme.

A key component of Eq.~\eqref{2ndordercorr} are the transition dipole
matrix elements between the electronic continuum states, $\langle
\phi_{k}^{-}|\hat{r}|\phi_{k_1}\rangle$.~Calculating these transition matrix
elements is challenging because of the delocalized nature of the continuum
states. Several approaches address this problem, for example
regularization~\cite{douguet2018PRA} or a hydrogenic approximation, which allows the matrix elements to be evaluated
analytically~\cite{dahlstrom2013ChemPhys, boll2022PRA, boll2023PRA}. We exploit the fact that the perturbed ionized states are semi-localized. The extent of their wavefunction increases incrementally with each application of the interaction operator. Therefore, for low-order perturbations, the perturbed state can still be represented by localized basis functions \cite{goetz2025PRR,Blech2025}. Accordingly, we represent the intermediate continuum states using unoccupied Hartree-Fock orbitals obtained from Molpro~\cite{werner2012molpro, werner2012WIRCMS}. We then replace the integral over $|k_1\rangle$ in Eq.~\eqref{2ndordercorr} with a summation over unoccupied Hartree–Fock orbitals,

\begin{equation}
	\begin{split}
		&\alpha^{k(2)}_{i_0}(t)
		= -\langle \phi_{k}^{-} | \hat{r} | \phi_{i_0} \rangle
		\sum_{i\in \mathrm{occ}} \langle \phi_i | \hat{r} | \phi_i \rangle
		\int_{-\infty}^{t} dt'\, e^{i(\varepsilon_{k} - \varepsilon_{i_0})t'} E(t')\\
		&\times\int_{-\infty}^{t'} dt''\, E(t'')
		- \sum_{b \notin \mathrm{occ}} \sum_{i\in \mathrm{occ}}
		\langle \phi_{k}^{-} | \hat{r} | \phi_{b} \rangle
		\langle \phi_{b} | \hat{r} | \phi_{i} \rangle\\
		&\times\int_{-\infty}^{t} dt'\, e^{i(\varepsilon_{k} - \varepsilon_{b})t'} E(t')
		\int_{-\infty}^{t'} dt''\,
		e^{i(\varepsilon_{b} - \varepsilon_{i})t''} E(t'').
	\end{split}
	\raisetag{1.5\baselineskip}
	\label{2ndordercorr(b)}
\end{equation}
\revB{The diagonal matrix elements $\langle \phi_i|\hat{r}|\phi_i\rangle$ appearing in
Eqs.~\eqref{2ndordercorr} and~\eqref{2ndordercorr(b)}, summed over occupied
orbitals, give the permanent dipole moment of the molecule. For methyloxirane,
as for any polar molecule, they do not vanish.

Transforming Eq.~\eqref{2ndordercorr(b)} into the laboratory frame,
substituting it into Eq.~\eqref{PAD1ph2ph}, and comparing the result with
Eq.~\eqref{padeq}, we obtain the orientation-averaged laboratory-frame
two-photon anisotropy parameters,
\begin{equation}
	\begin{split}
		&\beta_{L,M}^{(2\mathrm{ph})}(k)
		= (-1)^{M}\sum_{\mu_{0},\mu_{0}'}\sum_{\nu_{0},\nu_{0}'}(-1)^{\mu_{0}+\nu_{0}}
		\sum_{\mu,\mu'}\sum_{\nu,\nu'}(-1)^{\mu+\nu}\\
		&\times\sum_{l,m}\sum_{l',m'}(-1)^{m'}\,\eta(l,l',L) 
		\sum_{p,p'\geq i_{0}}
		F_{\mu_{0},\nu_{0}}^{(2)\,p,i_{0}}(k,\infty)\, \\
		&\times F_{\mu_{0}',\nu_{0}'}^{(2)\,p',i_{0}}(k,\infty)^{*}\,
		I_{l,m,\mu,\nu}^{(2)\,p,i_{0}}(k)\,
		I_{l',m',\mu',\nu'}^{(2)\,p',i_{0}}(k)^{*} \\
		&\times\sum_{j,j'=0}^{2}
		g_{\mu,\nu,\mu_{0},\nu_{0}}^{j}\,
		g_{\mu',\nu',\mu_{0}',\nu_{0}'}^{j'}
		\begin{pmatrix} l & l' & L \\ 0 & 0 & 0 \end{pmatrix} \begin{pmatrix} l & l' & L \\ m & -m' & m'-m \end{pmatrix} \\
		&\times\begin{pmatrix} j & j' & L \\ \mu+\nu & -\mu'-\nu' & m'-m \end{pmatrix} \begin{pmatrix} j & j' & L \\ -\mu_{0}-\nu_{0} & \mu_{0}'+\nu_{0}' & -M \end{pmatrix},
	\end{split}
	\label{eq:beta2ph}
\end{equation}
where, $F^{(2)\,p,i_0}_{\mu_0,\nu_0}(k,\infty)$ and
$I^{(2)\,p,i_0}_{l,m,\mu,\nu}(k)$ are the second-order analogs of the field
and dipole factors in Eq.~\eqref{eq:defs1ph} and the coefficients
\begin{equation*}
	g^{j}_{\mu,\nu,\mu_0,\nu_0} \equiv (2j+1)
	\begin{pmatrix} 1 & 1 & j \\ \mu & \nu & -\mu-\nu \end{pmatrix}
	\begin{pmatrix} 1 & 1 & j \\ -\mu_0 & -\nu_0 & \mu_0+\nu_0 \end{pmatrix}.
	\label{eq:gj}
\end{equation*}}
Explicit expressions and detailed derivations of all above parameter are given in Refs.~\cite{goetz2019JCP, goetz2019PRL, Blech2025}. 
The one- and two-photon contributions span different ranges of $L$: the
one-photon anisotropy parameters $\beta_{LM}^{(1\mathrm{ph})}$ are nonzero
only for $L \le 2$ and the two-photon anisotropy parameters
$\beta_{LM}^{(2\mathrm{ph})}$ only for $L \le 4$ \cite{goetz2019PRL}.
Table~\ref{beta_table} in Appendix~\ref{app:betas} lists all nonzero anisotropy parameters for the case of
circularly polarized XUV and linearly polarized IR and indicates their
contributions from the one-photon and two-photon pathways. \lg{Since this table
is given in the appendix, perhaps this reference should be moved there as well?
This needs to be somehow smoother.}
Among these anisotropy parameters, $\beta_{00}$ corresponds to the photoelectron spectrum (PES), which represents the total ionization yield. The PES is obtained by integrating Eq.~\eqref{padeq} over the solid angle $\Omega_k$, yielding
\begin{equation}
\mathrm{PES}(\epsilon_{k}) = \sqrt{4(\pi)}\,\beta_{00}^{\pm}(\epsilon_{k}) .
	\label{PES}
\end{equation}

We can express the PECD in terms of anisotropy parameters. For harmonics generated by one-photon ionization with circularly polarized XUV light, only $M=0$ terms contribute. In this case, PECD takes the form
\begin{equation}
	\mathrm{PECD}_{H}(\epsilon_{k},\theta_k)
	= \beta_{10}^{(1\mathrm{ph})}(\epsilon_{k})
	\sqrt{\frac{3}{\pi}} \cos \theta_k 
	\label{pecd_H}
\end{equation}
At the sidebands, where two-photon XUV+IR pathways contribute, PECD depends on additional angular components,
\raggedbottom
\begin{equation}
	\mathrm{PECD}_{S}(\epsilon_{k},\theta_k,\phi_{k})
	= \sum_{L=1}^{3} \sum_{M=0,\pm2}
	2\beta_{LM}^{(2\mathrm{ph})}(\epsilon_{k})
	Y_M^{L}(\theta_k,\phi_{k}) .
	\label{pecd_S}
\end{equation}

\section{\label{SecAugment}Basis Set Augmentation Scheme}
\begin{table}[tbp]
	\caption{Basis sets used in this study. Column 1 lists the basis set names and their acronyms. Columns 2--4 give the number of diffuse $s$-, $p$-, and $d$-type Gaussian basis functions added to the aug-cc-pVQZ basis set at the molecular charge center. The last column reports the total number of Gaussian functions introduced at the charge center. Each $s$-, $p$-, and $d$-type basis function contributes 1, 3, and 5 Gaussian functions, respectively. \textit{Note:} The counts include only the diffuse functions added at the molecular charge center.}
	\label{tab:basis_sets}
	\begin{ruledtabular}
		\begin{tabular}{lcccc}
			Basis set (short form) & $s$-type & $p$-type & $d$-type & Total \\
			\colrule
			cc-pVQZ (vqz)         & 0 & 0 & 0 & 0 \\
			aug-cc-pVQZ (avqz)    & 0 & 0 & 0 & 0 \\
			aug-cc-pVQZ-d (avqz-d)   & 1 & 1 & 1 & 9 \\
			aug-cc-pVQZ-2d (avqz-2d) & 2 & 2 & 2 & 18 \\
			aug-cc-pVQZ-4d (avqz-4d) & 4 & 4 & 4 & 36 \\
			aug-cc-pVQZ-6d (avqz-6d) & 6 & 6 & 6 & 54 \\
			aug-cc-pVQZ-8d (avqz-8d) & 8 & 8 & 8 & 72 \\
		\end{tabular}
	\end{ruledtabular}
\end{table}
The success of the approximation made in Eq.~\eqref{2ndordercorr(b)} depends on the convergence of the Gaussian basis sets used to obtain the Hartree-Fock orbitals. To address this convergence, we improve the
representation of the intermediate continuum states by refining the Gaussian
basis set.  We achieve this by
augmenting the aug-cc-pVQZ basis set with additional diffuse functions.
The augmentation of cc-pVXZ (X = D, T, Q,
\ldots)~basis sets with diffuse functions is well established and was
originally introduced to improve the description of
anions~\cite{kendall1992JCP}. We adopt the augmentation scheme of Ref.~\cite{goetz2017JCP} and
extend it by increasing the number of diffuse functions to improve the
representation of intermediate continuum states within the time-dependent perturbation theory
framework. To avoid linear-dependency issues associated with highly augmented
Gaussian basis sets, we place the additional diffuse functions at the molecular
charge center. We add primitive diffuse $s$-, $p$-, and $d$-type Gaussian basis
functions with the smallest exponent set to 0.015 $a_0^{-2}$. The remaining diffuse
functions were generated in a geometric sequence with a ratio of approximately
0.33, consistent with the ratio between the two smallest exponents in the
aug-cc-pVQZ basis set~\cite{kendall1992JCP}.

We use seven different basis sets, each containing a different number of diffuse functions,~cc-pVQZ, aug-cc-pVQZ, and aug-cc-pVQZ-\emph{n}d with \emph{n} = 1, 2, 4, 6, and 8, where the suffix ``d'' denotes diffuse functions (not the $d$-type basis function). For the remainder of the manuscript, we refer to them using the abbreviated names vqz, avqz, and avqz-\emph{n}d, respectively. The vqz basis set is the standard correlation-consistent quadruple-zeta basis set, while avqz is its augmented version that includes one additional diffuse basis function on each atom. The avqz-\emph{n}d basis sets are constructed by adding \emph{n} diffuse basis functions of each angular-momentum type ($s$, $p$, and $d$) to the avqz basis set. Details of these basis sets are provided in Table~\ref{tab:basis_sets}. \revA{Table~\ref{orbeng:table} lists the energies of the HOMO, HOMO$-1$, the lowest unoccupied molecular orbital
(LUMO), and LUMO$+1$. The occupied orbital energies are already converged at the avqz level, whereas the unoccupied orbital energies approach zero as increasingly diffuse functions are added. In our HF/avqz-\emph{n}d calculations, the HOMO is delocalized over the epoxide ring and is predominantly assigned as $\sigma_\mathrm{CO}$, while the HOMO$-1$ is the oxygen lone pair ($n_\mathrm{O}$); the HOMO$-2$ and HOMO$-3$ are of $\sigma_\mathrm{CH}/n_\mathrm{O}$ and $\sigma_\mathrm{CC}$ character, respectively. The latter two agree with Ref.~\cite{MendesPCCP2026}, whereas the characters of the two highest orbitals appear in reverse order. Within Koopmans' theorem, the HOMO energy gives an estimated ionization energy of 11.82\,eV, which overestimates the experimental values. In the present calculations, we therefore adopt an ionization energy $|\epsilon_0| = 10.35$\,eV, consistent with the theoretical calculations of Ref.~\cite{han2025Nature} and within the range of experimentally reported first ionization energies of 9.80--10.44\,eV~\cite{gallegos1961JACS,WATANABE1962JQSRT,McAlduff1977CJC,kimura1981JSSP,liu1999JPCA,rodrigues2024JPCA}.}

\begin{table}[tbp]
	\revA{\caption{Hartree-Fock orbital energies (in eV) of the two highest occupied
			(HOMO$-1$, HOMO) and two lowest unoccupied (LUMO, LUMO$+1$) molecular
			orbitals of methyloxirane for different basis sets. The occupied orbitals (HOMO and HOMO-1)
			energies are converged already at the avqz level, whereas the unoccupied
			orbital (LUMO and LUMO+1) energies approach the ionization threshold as diffuse functions
			are added.}
	\label{orbeng:table}
	\begin{ruledtabular}
		\begin{tabular}{lcccc}
			Basis Sets & HOMO$-1$ & HOMO & LUMO & LUMO$+1$ \\
			\colrule
			vqz     & $-12.1232$ & $-11.8114$ & $3.1604$ & $3.8706$ \\
			avqz    & $-12.1366$ & $-11.8231$ & $0.7026$ & $0.9083$ \\
			avqz-d  & $-12.1366$ & $-11.8232$ & $0.4294$ & $0.6422$ \\
			avqz-2d & $-12.1366$ & $-11.8232$ & $0.1633$ & $0.2683$ \\
			avqz-4d & $-12.1366$ & $-11.8232$ & $0.0144$ & $0.0321$ \\
			avqz-6d & $-12.1366$ & $-11.8232$ & $0.0012$ & $0.0035$ \\
			avqz-8d & $-12.1366$ & $-11.8232$ & $0.0001$ & $0.0003$ \\
		
		\end{tabular}
	\end{ruledtabular}
}
\end{table}
We test the convergence of the photoelectron circular dichroism (PECD) [Eq.~\eqref{pecdeq}], photoelectron spectrum (PES) [Eq.~\eqref{PES}], and anisotropy parameters ($\beta_{L,M}$) [Eq.~\eqref{padeq}] with respect to the additional diffuse functions listed in Table~\eqref{tab:basis_sets}. We isolate the role of these diffuse functions by keeping fixed the truncation of the unoccupied Hartree–Fock orbitals as defined in Eq.~\eqref{2ndordercorr(b)}. In all calculations, we include only unoccupied Hartree–Fock orbitals with energies up to 12$\,$eV. We choose this cutoff because we consider harmonics H\,9–H\,13, for which the highest accessible sideband energy is 11.35$\,$eV (SB\,14), ensuring that all energetically relevant intermediate states are included while excluding higher-lying virtual orbitals.

\section{\label{sec3}Results and Discussions} 
We model the XUV field as a circularly polarized frequency comb consisting of
the harmonics $\omega_{7}$, $\omega_{9}$, $\omega_{11}$, and $\omega_{13}$ of
a 1.55\,eV fundamental frequency. \revA{All harmonic components have a Gaussian
intensity envelope with a FWHM of 5\,fs. In the
time domain, their coherent superposition forms a train of attosecond bursts
separated by half the fundamental period ($\approx 1.33$\,fs); the train
contains three to four bursts within its FWHM, each with a duration of
$\approx 300$\,as for equal harmonic phases. The individual bursts are
therefore in the attosecond regime, while the 5\,fs envelope sets the overall
length of the train.} \revB{ Each harmonic component has a peak intensity of
$\approx 2.2\times10^{11}$\,W/cm$^2$, and the IR field is taken to be about
twice as intense, $\approx 4.4\times10^{11}$\,W/cm$^2$. These intensities are
consistent with the theoretical calculations of Ref.~\cite{han2025Nature},
where they were chosen to best reproduce the widths and relative intensities
of the peaks in the experimentally measured photoelectron spectrum.} For the convergence analysis of PECD and $\beta_{L,M}$, we use an IR pulse with a 15$\,$fs FWHM, linearly polarized along the x-axis. For the convergence of the forward--backward delay, we adopt the same IR configuration (a linearly polarized pulse with a 5$\,$fs FWHM) as in Ref.~\cite{han2025Nature}, which allows a direct comparison with the forward--backward delay results reported therein. We also repeat the convergence analysis for PECD and $\beta_{L,M}$ using the IR configuration from
Ref.~\cite{han2025Nature} and observe the same convergence trends across
different laser configurations (see Appendix). We select a photoelectron energy
range of 0.5--12\,eV, motivated by previous
studies~\cite{goetz2019PRL,goetz2025PRR,han2025Nature}, to evaluate the
convergence. Specifically, we examine harmonics H\,7, H\,9, H\,11, and H\,13, which
correspond to photoelectron energies of 0.5$\,$eV, 3.6$\,$eV, 6.7$\,$eV, and 9.8$\,$eV,
respectively, as well as the associated sidebands SB\,8, SB\,10, SB\,12, and SB\,14,
corresponding to photoelectron energies of 2.05$\,$eV, 5.15$\,$eV, 8.25$\,$eV, and
11.35$\,$eV, respectively.

\begin{figure}[htbp]
	\centering
	\includegraphics[width=\columnwidth]{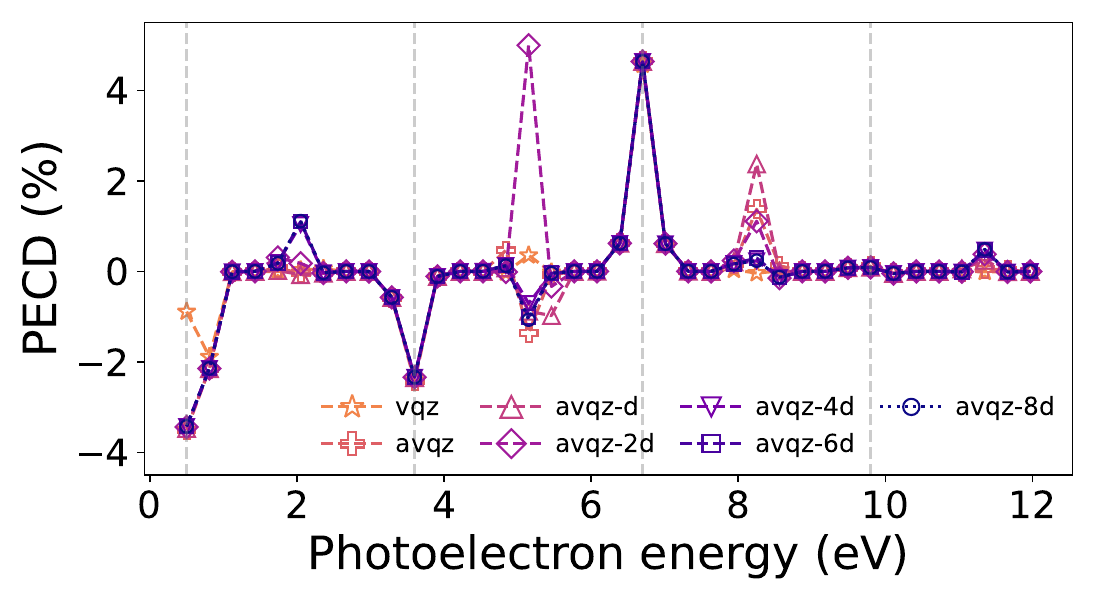}
	\caption{The PECD at $\theta=\phi=0^\circ$ and $\tau_k=0$ is shown as a function of photoelectron energy, normalized to the maximum photoelectron intensity. The vertical gray dashed lines indicate the harmonic positions. The PECD at the harmonics (one-photon ionization) shows rapid convergence, whereas convergence at the sidebands (two-photon ionization) requires the inclusion of additional diffuse functions.}	
	\label{PECDplot}
\end{figure}

\begin{table*}[t]
	\caption{Numerical values of PECD at the harmonics (H 7--H 13) and sidebands (SB 8--SB 14), corresponding to Fig.~\protect\ref{PECDplot}. The table quantifies the convergence of the PECD with respect to the diffuse functions; converged digits are highlighted in red. The single-photon PECD converges rapidly to at least four digits, whereas the two-photon PECD converges more slowly. In particular, SB10 shows no converged digits for the PECD evaluated at $\theta=\phi=0^\circ$ and $\tau_k=0$.}
	\label{pecd table}
	\begin{ruledtabular}
		\begin{tabular}{lcccccccc}
			& H7 & SB8 & H9 & SB10 & H11 & SB12 & H13 & SB14 \\
			\colrule
			vqz     & -0.88800 & 0.03065 & \textcolor{red}{-2.}34946 & 0.36056 & \textcolor{red}{4.}56873 & -0.02254 & \textcolor{red}{0.}10851 & -0.00827 \\
			avqz    & \textcolor{red}{-3.}50329 & -0.03586 & \textcolor{red}{-2.}42826 & -1.36046 & \textcolor{red}{4.6}5353 & 1.35864 & \textcolor{red}{0.}11376 & \textcolor{red}{0.}05475 \\
			avqz-d  & \textcolor{red}{-3.4}7175 & -0.07056 & \textcolor{red}{-2.3}4700 & -0.97722 & \textcolor{red}{4.6}3827 & 2.32803 & \textcolor{red}{0.}09239 & \textcolor{red}{0.}27769 \\
			avqz-2d & \textcolor{red}{-3.4}3641 & 0.17976 & \textcolor{red}{-2.33}762 & 4.94759 & \textcolor{red}{4.638}89 & 1.10879 & \textcolor{red}{0.08}756 & \textcolor{red}{0.}45726 \\
			avqz-4d & \textcolor{red}{-3.42}526 & \textcolor{red}{1.}04149 & \textcolor{red}{-2.335}36 & -0.68550 & \textcolor{red}{4.6387}6 & \textcolor{red}{0.3}1270 & \textcolor{red}{0.086}18 & \textcolor{red}{0.515}58 \\
			avqz-6d & \textcolor{red}{-3.424}32 & \textcolor{red}{1.1}0234 & \textcolor{red}{-2.3352}3 & -0.99010 & \textcolor{red}{4.6387}4 & \textcolor{red}{0.32}042 & \textcolor{red}{0.086}10 & \textcolor{red}{0.515}34 \\
			avqz-8d & \textcolor{red}{-3.424}27 & \textcolor{red}{1.1}2585 & \textcolor{red}{-2.3352}0 & -1.02102 & \textcolor{red}{4.6387}6 & \textcolor{red}{0.32}195 & \textcolor{red}{0.086}09 & \textcolor{red}{0.515}26 \\
		\end{tabular}
	\end{ruledtabular}
\end{table*}
\subsection{Convergence of the PECD}
We first examine the convergence of the PECD. Figure~\ref{PECDplot} (and Table~\ref{pecd table}) shows the PECD evaluated at zero XUV–IR delay ($\tau_k = 0$), and photoelectron emission angles of $\theta = \phi = 0^{\circ}$, for the basis sets listed in Table~\ref{tab:basis_sets}. Table~\ref{pecd table} presents the numerical values of the PECD shown in Fig.~\ref{PECDplot} at the harmonics (H\,7--H\,13) and sidebands (SB\,8--SB\,14).
\begin{figure}[tbp]
	\includegraphics[width=\columnwidth]{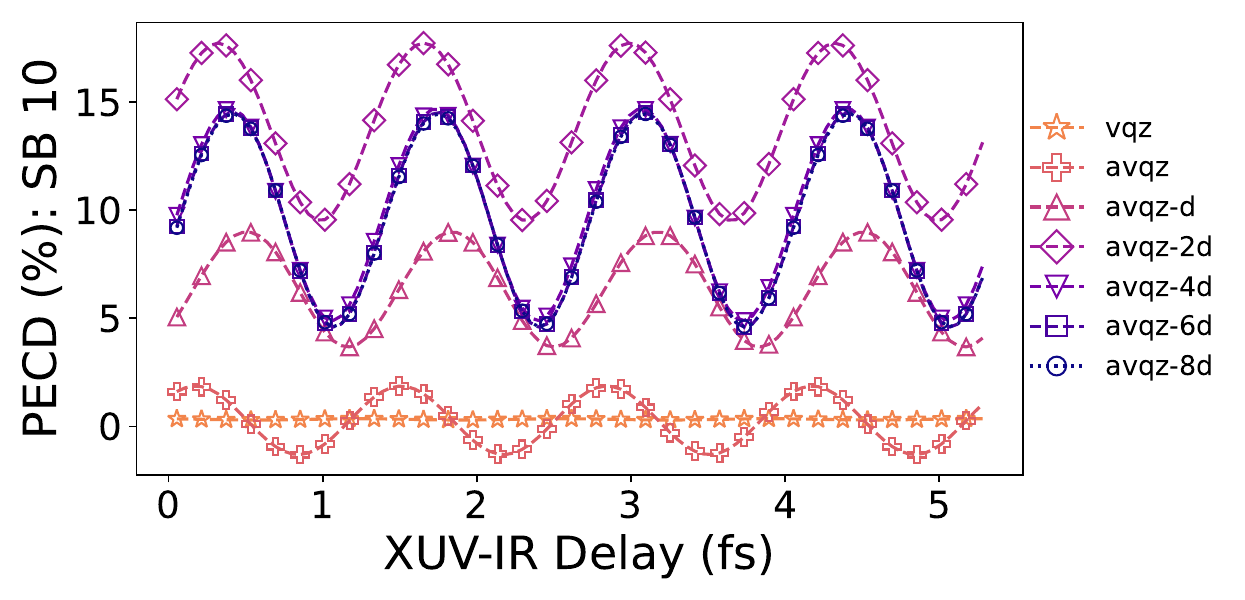}
	\caption{PECD as a function of the XUV--IR delay for SB 10 at a photoelectron energy of 5.15$\,$eV. The peak PECD converges to 14.5\% as diffuse functions are added. The PECD at SB 10 is one of the most sensitive observables in this study; neglecting diffuse functions underestimates the PECD by nearly an order of magnitude.}
	\label{pecdtimedelay}
\end{figure}
The PECD at the harmonic energies, corresponding to single-photon ionization [Eq.~\eqref{pecd_H}], converges rapidly. This trend is consistent with Ref.~\cite{goetz2025PRR}, which also finds that the initial electronic states are well converged with respect to the Gaussian basis sets. This result is expected, since within the time-dependent perturbation theory framework, the single-photon response involves only the initial electronic states and does not require the unoccupied Hartree--Fock orbitals (see Eq.~\eqref{1stordercorr}). The only effect of the basis set on the convergence of the harmonics arises from its influence on the ground state. Adding more diffuse functions changes the ground-state orbitals. These changes lead to corresponding modifications of the Coulomb and exchange potential operators used in the scattering calculations~\cite{Lucchese1982PRA,natalense1999JCP}. As shown in  Fig.~\ref{PECDplot} (and Table~\ref{pecd table}), the PECD at H\,9 and H\,11 converges to four decimal places, while at H\,7 and H\,13 it converges to three decimal places. In contrast, PECD at the sidebands, which involves two-photon ionization, converges at a slower rate. With increasing numbers of diffuse functions in the basis set, the PECD converges to two digits for SB\,8, three digits for SB\,12, and four digits for SB\,14, while SB\,10 does not converge to a stable digit and exhibits the slowest convergence among the sidebands. This overall slower convergence at sidebands mainly arises because the PECD at the sidebands depends on intermediate continuum states, which are only semi-localized rather than fully bound within the time-dependent perturbation theory framework, and therefore require more diffuse Gaussian functions for accurate representation. Our augmentation approach, described above, systematically increases the number of unoccupied Hartree--Fock orbitals and thereby improves the convergence of the PECD at sidebands with respect to the added diffuse functions.
\begin{table*}[t]
	\caption{$\beta_{00}$ values at the harmonics (H\,7–H\,13) and sidebands (SB\,8–SB\,14). The table quantifies the convergence of the $\beta_{00}$ (PES) with respect to the diffuse functions; converged digits are highlighted in red. At the harmonics, $\beta_{00}$ (PES) exhibits rapid and monotonic convergence, whereas at the sidebands it displays erratic, non-monotonic behavior (e.g., SB\,8 and SB\,10), requiring extensive diffuse functions (up to avqz-6d) for convergence. Notably, PES converges faster than PECD (Table~\protect\ref{pecd table}); even the difficult SB\,10 case demonstrates faster convergence.}
	\label{padtable}
	\begin{ruledtabular}
		\begin{tabular}{lcccccccc}
			& H7 & SB8 & H9 & SB10 & H11 & SB12 & H13 & SB14 \\
			\colrule
			vqz & \textcolor{red}{0.8}49675 & \textcolor{red}{0.0}03093 & \textcolor{red}{0.404}378 & \textcolor{red}{0.}036688 & \textcolor{red}{0.38}3851 & \textcolor{red}{0.}019074 & \textcolor{red}{0.2}90193 & \textcolor{red}{0.}001391 \\
			avqz & \textcolor{red}{0.86}1195 & \textcolor{red}{0.0}14804 & \textcolor{red}{0.404}627 & \textcolor{red}{0.}139869 & \textcolor{red}{0.38}7120 & \textcolor{red}{0.}076763 & \textcolor{red}{0.286}599 & \textcolor{red}{0.}010484 \\
			avqz-d & \textcolor{red}{0.86}1948 & \textcolor{red}{0.0}11673 & \textcolor{red}{0.404}209 & \textcolor{red}{0.}679870 & \textcolor{red}{0.3869}89 & \textcolor{red}{0.}337969 & \textcolor{red}{0.286}881 & \textcolor{red}{0.}037520 \\
			avqz-2d & \textcolor{red}{0.862}568 & \textcolor{red}{0.0}46212 & \textcolor{red}{0.4040}68 & \textcolor{red}{0.}707937 & \textcolor{red}{0.3869}60 & \textcolor{red}{0.}383988 & \textcolor{red}{0.286}953 & \textcolor{red}{0.}074057 \\
			avqz-4d & \textcolor{red}{0.8629}13 & \textcolor{red}{0.08}3507 & \textcolor{red}{0.40402}6 & \textcolor{red}{0.62}3316 & \textcolor{red}{0.3869}49 & \textcolor{red}{0.404}261 & \textcolor{red}{0.28697}3 & \textcolor{red}{0.1017}82 \\
			avqz-6d & \textcolor{red}{0.8629}39 & \textcolor{red}{0.085}630 & \textcolor{red}{0.404023} & \textcolor{red}{0.6201}80 & \textcolor{red}{0.3869}49 & \textcolor{red}{0.4047}01 & \textcolor{red}{0.28697}5 & \textcolor{red}{0.1017}21 \\
			avqz-8d & \textcolor{red}{0.8629}41 & \textcolor{red}{0.085}436 & \textcolor{red}{0.404023} & \textcolor{red}{0.6201}30 & \textcolor{red}{0.3869}50 & \textcolor{red}{0.4047}95 & \textcolor{red}{0.28697}7 & \textcolor{red}{0.1017}14 \\
		\end{tabular}
	\end{ruledtabular}
\end{table*}

\begin{figure*}[tbp]
	\centering
	\includegraphics[width=\textwidth]{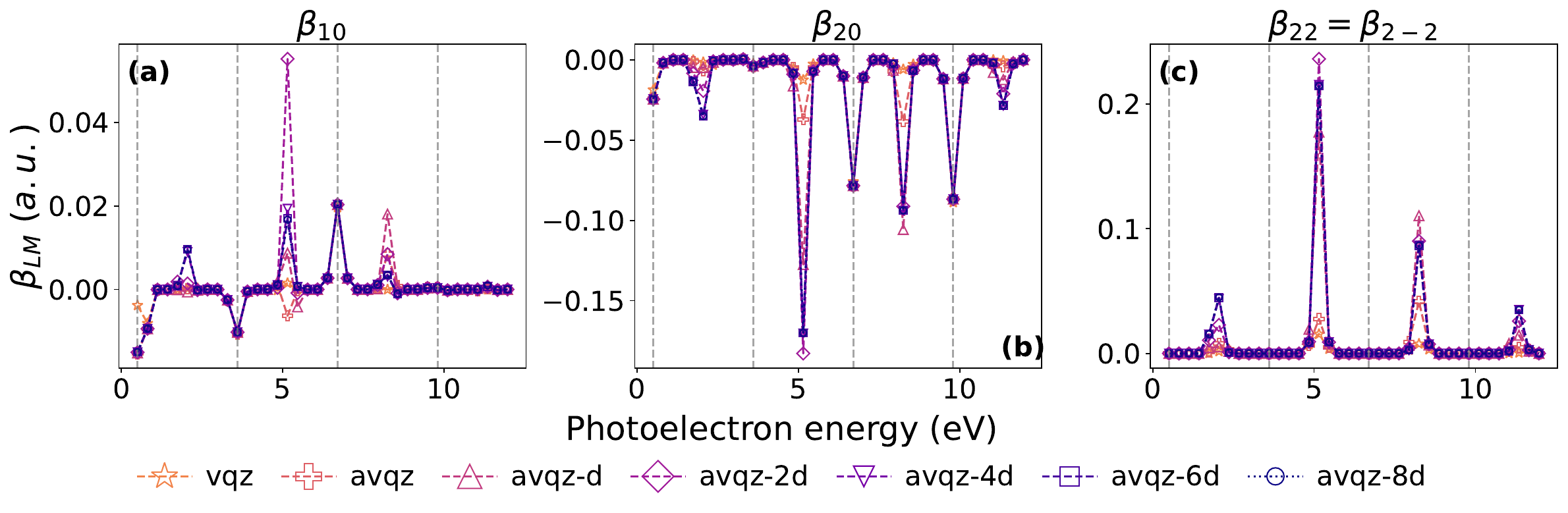}
	
	\caption{Anisotropy parameters ($\beta_{LM}$) plotted as functions of photoelectron energy: (a) $\beta_{10}$ $(L=1,M=0)$, (b) $\beta_{20}$ $(L=2,M=0)$, and (c) $\beta_{22}$ $(L=2,M=2)$. These parameters receive contributions from both one- and two-photon ionization pathways. Vertical gray dashed lines mark the harmonic positions. The figure illustrates the convergence of these parameters with respect to the inclusion of diffuse functions. Compared with $\beta_{00}$ (Table~\protect\ref{padtable}), they converge more slowly at both harmonics and sidebands. As higher-order anisotropy parameters ($L\ge1$), they exhibit greater convergence difficulty and require more diffuse functions than $\beta_{00}$.}
	
	\label{betas}
\end{figure*}
We next examine the convergence of the delay-resolved PECD for nonzero XUV-IR delays. At all sidebands, the PECD exhibits a periodic modulation with a period of 1.33$\,$fs, corresponding to $2\omega_{\mathrm{IR}}$, consistent with conventional RABBITT measurements~\cite{han2025Nature}. Figure~\ref{pecdtimedelay} shows the delay-dependent PECD for SB\,10, corresponding to the photoelectron energy of 5.15$\,$eV. We focus on SB\,10 because it represents the most challenging case, with the slowest convergence (see Table~\ref{pecd table}) and the largest modulation amplitude among all sidebands. As shown in Fig.~\ref{pecdtimedelay}, the peak PECD at SB\,10 converges to two digits, reaching a value of approximately 14\% (see Table~\ref{delaypecd:table} of Appendix~\ref{app:pecd}). The other sidebands exhibit similar behavior, converging to values consistent within at least one decimal place (Fig.~\ref{delayed PECD SB} and Table~\ref{delaypecd:table}).

The convergence of the PECD reflects more than the observable itself: what converges are the two-photon transition amplitudes involving intermediate continuum states, accurately represented by unoccupied Hartree--Fock orbitals in a sufficiently augmented Gaussian basis. 
Since the same amplitudes govern multiphoton ionization in molecules in general, the Gaussian-based time-dependent perturbation theory framework should be applicable well beyond the present chiral observable, while the converged PECD itself provides the quantitative accuracy required for the coherent control of the PECD~\cite{goetz2019PRL,goetz2025PRR}.

\subsection{Convergence of the Anisotropy Parameters}

Next, we examine the convergence of the real parts of all nonzero anisotropy parameters with respect to the diffuse basis sets listed in Table~\ref{tab:basis_sets}; the imaginary parts follow qualitatively the same convergence behavior through the symmetry relations of Table~\ref{beta_table} and are therefore not shown separately. For the parameters up to $L=2$, $\beta_{00}$ is given in Table~\ref{padtable}, while $\beta_{10}$, $\beta_{20}$, and
$\beta_{22}=\beta_{2-2}$ are shown in Fig.~\ref{betas}. The higher-order parameters $(L=3,4)$ are reported in Tables~\ref{tab:beta3_conv} and \ref{tab:beta4_conv} of Appendix~\ref{app:betas}, which also presents a test of the dependence on the IR pulse duration. Of these, $\beta_{00}$, $\beta_{10}$, and
$\beta_{20}$ receive contributions from both one- and two-photon pathways,
whereas the remaining ones arise solely from two-photon pathways
(Table~\ref{beta_table}).

The parameter $\beta_{00}$ is proportional to the PES (see Eq.~\eqref{PES}). Table~\ref{padtable} lists its values for the harmonics H\,7–H\,13 and sidebands SB\,8–SB\,14, with converged values highlighted in red. It converges rapidly at the harmonics: to the sixth decimal place for H\,9, the fourth for H\,7 and H\,11, and the fifth for H\,13. The sidebands converge more slowly, with SB\,10, SB\,12, and SB\,14 converging to four decimal places and SB\,8 to three. The origin of this convergence behavior is the same as for PECD discussed above.

The parameter $\beta_{10}$, shown in Fig.~\ref{betas}(a), gives the PECD at
the harmonics through its one-photon component [Eq.~\eqref{pecd_H}] and
contributes to the PECD at the sidebands through its two-photon component
[Eq.~\eqref{pecd_S}]. It converges to more than five decimal places at the
harmonics, while at the sidebands convergence is slower: three decimal places
at SB\,8, SB\,12, and SB\,14, and two at SB\,10. The parameters $\beta_{20}$
and $\beta_{22}=\beta_{2-2}$, shown in Figs.~\ref{betas}(b) and \ref{betas}(c),
also contribute to the PECD at the sidebands [Eq.~\eqref{pecd_S}]. Like
$\beta_{10}$, the parameter $\beta_{20}$ converges to more than five decimal
places at the harmonics, but shows uneven convergence at the sidebands: two
decimal places at SB\,8, one at SB\,10, and four at SB\,12 and SB\,14.
$\beta_{22}$ converges to three decimal places at SB\,8, SB\,10, and SB\,12,
and to four at SB\,14. The higher-order parameters converge similarly at the
sidebands: $\beta_{30}$ and $\beta_{32}$ (Table~\ref{tab:beta3_conv}), which likewise contribute to the
PECD at the sidebands [Eq.~\eqref{pecd_S}], converge to three decimal places
at all sidebands, except $\beta_{32}$, which reaches four decimal places at SB\,10;
$\beta_{40}$ converges to four decimal places at SB\,8, SB\,12, and SB\,14,
and three at SB\,10, while $\beta_{42}$ converges to four decimal places at
SB\,8, three at SB\,10 and SB\,12, and five at SB\,14 (Table~\ref{tab:beta4_conv}).
\begin{figure}[t]
	\includegraphics[width=\columnwidth]{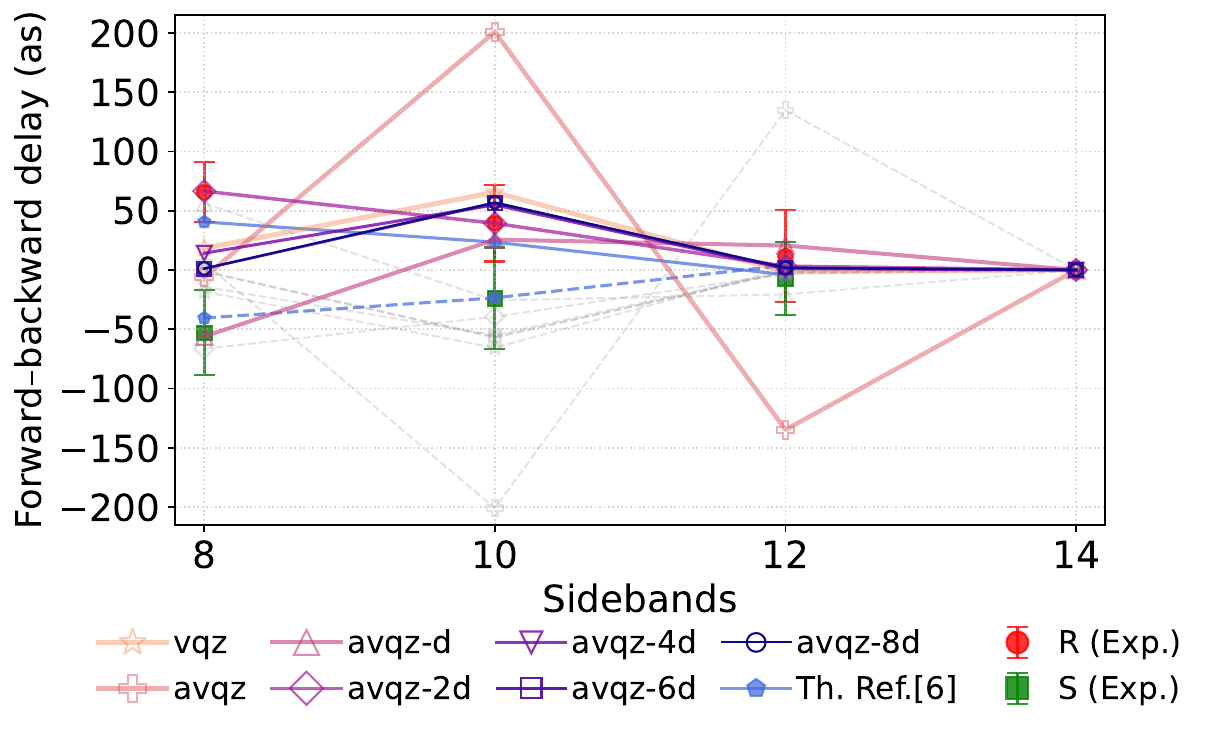}
	\caption{Forward–backward photoionization time delay at the sidebands. Solid and dashed (gray) lines denote the R- and S-type enantiomers, respectively. The present results (labeled vqz through avqz-8d) illustrate the convergence of the calculations with increasingly diffuse basis sets. The forward–backward time delay is highly sensitive and converges more slowly than the PECD (Table~\protect\ref{pecd table}) and PES (Table~\protect\ref{padtable}). We compare the converged results with theoretical predictions (blue lines) and experimental measurements (error bars) from Ref.~\cite{han2025Nature}, available at SB\,8, SB\,10, and SB\,12. The converged delays lie within the experimental uncertainties at SB\,10 and SB\,12, while the SB\,8 deviation likely arises from neglected electron correlation.}
	\label{fbdelay}
\end{figure}

Comparing the PECD values in Table~\ref{pecd table} with the $\beta_{00}$ (PES) values in Table~\ref{padtable} shows that $\beta_{00}$ converges significantly faster than the PECD. At the sidebands, the PECD converges to two to three decimal places, whereas $\beta_{00}$ converges to four to five. Even in the most challenging case (SB\,10), $\beta_{00}$ converges substantially faster than the PECD. Furthermore, the higher-order anisotropy parameters ($\beta_{LM}$) also exhibit faster convergence profiles than the PECD. At the harmonics, the PECD similarly converges more slowly than both $\beta_{00}$ and the other $\beta_{LM}$ parameters. One reason for this slower convergence may be that the PECD is a difference-based observable [see Eq.~\eqref{pecdeq}], which makes it inherently more sensitive and therefore slower to converge than the anisotropy parameters. 

The anisotropy parameters constitute the complete PAD, and its asymmetry is determined by the pulse polarization and the number of photons involved in the ionization process, which together determine the set of nonzero $\beta_{LM}$ (Table~\ref{beta_table})~\cite{Blech2025,goetz2025PRR}. Their convergence confirms that the Gaussian-based time-dependent perturbation theory framework accurately describes the photoelectron angular distributions in the multiphoton ionization of molecules---a prerequisite, in particular, for the coherent control of the PAD.

\begin{table}[tbp]
	\caption{Forward–backward photoionization time delay (as) for the R enantiomer at sidebands SB 8–SB 14, computed with increasingly diffuse basis sets (vqz through avqz-8d). The corresponding data are shown in Fig.~\ref{fbdelay}.}
	\label{fb:table}
	\begin{ruledtabular}
		\begin{tabular}{lcccc}
			Basis set & SB 8 & SB 10 & SB 12 & SB14 \\
			\colrule
			vqz     & 18.356  & 65.553  & -1.351   & -0.000 \\
			avqz    & -6.187  & 200.743 & -134.698 & 0.003 \\
			avqz-d  & -56.075 & 25.620  & 20.604   & 0.000 \\
			avqz-2d & 66.452  & 39.367  & 3.225    & -0.001 \\
			avqz-4d & 14.174  & 53.937  & 1.155    & -0.001 \\
			avqz-6d & 1.061   & 56.325  & 2.020    & -0.001 \\
			avqz-8d & 1.303   & 57.027  & 1.820    & -0.001 \\
		\end{tabular}
	\end{ruledtabular}
\end{table}

\subsection{Photoionization delays}

Reference~\cite{han2025Nature} reports a forward–backward chiral-sensitive delay of up to 60$\,$as for both enantiomers of C$_3$H$_6$O. Here, we revisit those calculations to examine the convergence of the forward–backward photoionization delay with respect to increasingly diffuse basis sets. Figure~\ref{fbdelay} shows these delays at each sideband, where the blue lines denote the theoretical results of Ref.~\cite{han2025Nature}, while the red and green error bars show the corresponding experimental values. Table~\ref{fb:table} lists the numerical values for the R-enantiomer at sidebands SB\,8–SB\,14; the S-enantiomer delays are equal in magnitude and opposite in sign. The forward–backward delay approaches convergence with respect to the inclusion of diffuse basis functions, with values stabilizing near 1$\,$as at SB\,8, 57$\,$as at SB\,10, and 1.8$\,$as at SB\,12 for the most diffuse basis sets, as shown in Table~\ref{fb:table}. At SB\,8, the current calculations converge to a value significantly smaller than both the theoretical results and the experimental data of Ref.~\cite{han2025Nature}. At SB\,10, the current result slightly exceeds the theoretical prediction of Ref.~\cite{han2025Nature} but remains within the experimental error bars. At SB\,12, the current calculations converge to values consistent with the experiment. The remaining discrepancies with respect to experiment likely reflect the neglect of electron–correlation effects in the present calculations; because correlation effects become stronger at lower photoelectron energies, their inclusion may be essential for quantitative agreement, particularly at SB\,8. 

We emphasize that the theoretical protocol of Ref.~\cite{han2025Nature} differs from the present one: there, the energies of the unoccupied Hartree--Fock orbitals are redistributed through a fitting procedure tuned to reproduce the experimental photoelectron spectrum and thereby suppress artificial resonances. This improves agreement with experiment but introduces an empirical parameter into the calculation. The present calculations, by contrast, follow a strictly \textit{ab initio} treatment without empirical fitting, so the agreement obtained at the most diffuse basis sets is a parameter-free result. The converged values have the experimental sign across all sidebands, whereas the fitted theory of Ref.~\cite{han2025Nature} reverses sign at SB\,12 (Fig.~\ref{fbdelay}).

\section{\label{sec5}Conclusions}

We have performed a rigorous convergence analysis of multiphoton PECD, photoelectron spectra, anisotropy parameters, and the
forward--backward photoionization time delay for randomly oriented chiral
molecules interacting with a linearly polarized IR field and a circularly
polarized XUV field, within a time-dependent perturbation theory framework. To address the challenge of evaluating transition dipole matrix
elements between electronic continuum states, the intermediate continuum
states are represented by unoccupied Hartree--Fock orbitals~\cite{goetz2019PRL, goetz2025PRR}. To improve this representation, we have augmented the standard aug-cc-pVQZ basis with additional diffuse functions placed at the molecular charge center.

The convergence analysis reveals two central results. First, the
theoretical framework of Ref.~\cite{goetz2019PRL} is confirmed to be
effective: all considered observables converge systematically with the
inclusion of diffuse basis functions. This is very gratifying, as it
confirms the adequacy of representing the intermediate continuum states
within the time-dependent perturbation theory framework by unoccupied Hartree--Fock orbitals in a
Gaussian basis sets, provided that a sufficient number of diffuse functions is
included. In particular, the augmentation resolves the basis set discrepancies in the two-photon PECD reported in Ref.~\cite{goetz2025PRR}. Second, the degree of diffuseness
required depends strongly on the observable: the photoelectron spectrum and the anisotropy parameter $\beta_{00}$ (which is proportional to photoelectron spectrum) converge with comparatively few
diffuse functions, whereas PECD requires significantly more. The two-photon PECD converges to two digits in most cases, while $\beta_{00}$ converges to three to four digits. All observables converge considerably faster in the one-photon regime. Furthermore, the convergence behavior of all observables is found to be independent of the IR pulse duration.

The forward--backward photoionization delay likewise converges systematically with the
number of diffuse functions. This observable constitutes the most stringent test of the present approach: as an enantiosensitive, attosecond-scale quantity, it is determined by the phases rather than the magnitudes of the two-photon transition amplitudes, such that even small deficiencies in the continuum representation manifest as spurious delays. Its convergence therefore establishes the reliability of the Gaussian-based time-dependent perturbation theory framework at the level of spectral phases, and not merely of ionization yields.

Obtained from fully \textit{ab initio} calculations, the converged
forward--backward photoionization delays are in qualitative agreement with the measurements of Ref.~\cite{han2025Nature} at the higher-energy sidebands, whereas previous theoretical descriptions of these data relied on empirical modeling. The remaining discrepancies, most pronounced at SB~8, likely originate from the neglect of electron correlation in the present
static-exchange description at the Hartree--Fock level. Such effects are
expected to be more significant at lower photoelectron energies. We are currently exploring the inclusion of these effects through a close-coupling treatment, which should improve the agreement at the lower sidebands.

The augmentation strategy established here is not specific to
methyloxirane, nor to chiral systems. The quantities converged in this work are the two-photon transition amplitudes involving continuum states, which underlie attosecond interferometric observables in general; PECD and the forward--backward photoionization delay serve here as particularly stringent, phase-sensitive tests. The same approach therefore applies directly to RABBITT and related interferometric schemes in achiral molecules, enabling
the \textit{ab initio} computation of angle-resolved photoionization time
delays. For chiral systems, it allows the fully \textit{ab initio} design
and analysis of coherent-control schemes, such as the optimization of pulse parameters to maximize the chiral response or the separation of bound- and continuum-state contributions to PECD~\cite{goetz2025PRR}. Combined with a close-coupling treatment of electron correlation, which we are currently exploring, the present framework should enable quantitative \textit{ab initio} simulations of attosecond photoionization dynamics in polyatomic molecules.

\section{\label{sec6}Acknowledgment}
We acknowledge R. Esteban Goetz for his contributions to the development of the TDPT code used in this work. The computing for this project was performed on the Beocat Research Cluster at Kansas State University, which is funded in part by NSF Grants No. CNS-1006860, No. EPS-1006860, No. EPS-0919443, No. ACI-1440548, No. CHE-1726332, and No. NIH P20GM113109, and used resources of the National Energy Research Scientific Computing Center (NERSC), a U.S. Department of Energy Office of Science User Facility operated under Contract No. DE-AC02-05CH11231 using NERSC Award No. BES-ERCAP0024357. M.S. and L.G. were supported by the Chemical Sciences, Geosciences, and Biosciences Division, Office of Basic Energy Sciences, Office  of Science, U.S. Department of Energy, under Grant No. DESC0022105. M.S. acknowledges support from the Dr. Shuo Zeng Scholarship for his research at Kansas State University and thanks Brett Esry for helpful discussions. A.B. and C.P.K. acknowledge financial support from the Deutsche Forschungsgemeinschaft (CRC 1319). The authors used AI-based language tools for grammar checking and sentence structure improvements.
\appendix

\section{PECD}
\renewcommand{\thefigure}{A\arabic{figure}}
\label{app:pecd}
\setcounter{figure}{0}
In the main text we establish PECD convergence using a 5$\,$fs (FWHM) XUV pulse and a 15$\,$fs IR pulse. To test the role of pulse duration, Fig.~\ref{pecd_theta0} shows the PECD versus photoelectron energy at fixed emission angle  ($\theta = 0^\circ$) and zero XUV–IR delay, computed with the same XUV field but a shorter 5$\,$fs (FWHM) IR pulse. Comparison with the 15$\,$fs case [Fig.~\ref{PECDplot}] shows that the convergence with respect to diffuse functions is unchanged, confirming that it is independent of the IR pulse duration.

\begin{figure}[b]
	\centering
	\includegraphics[width=\columnwidth]{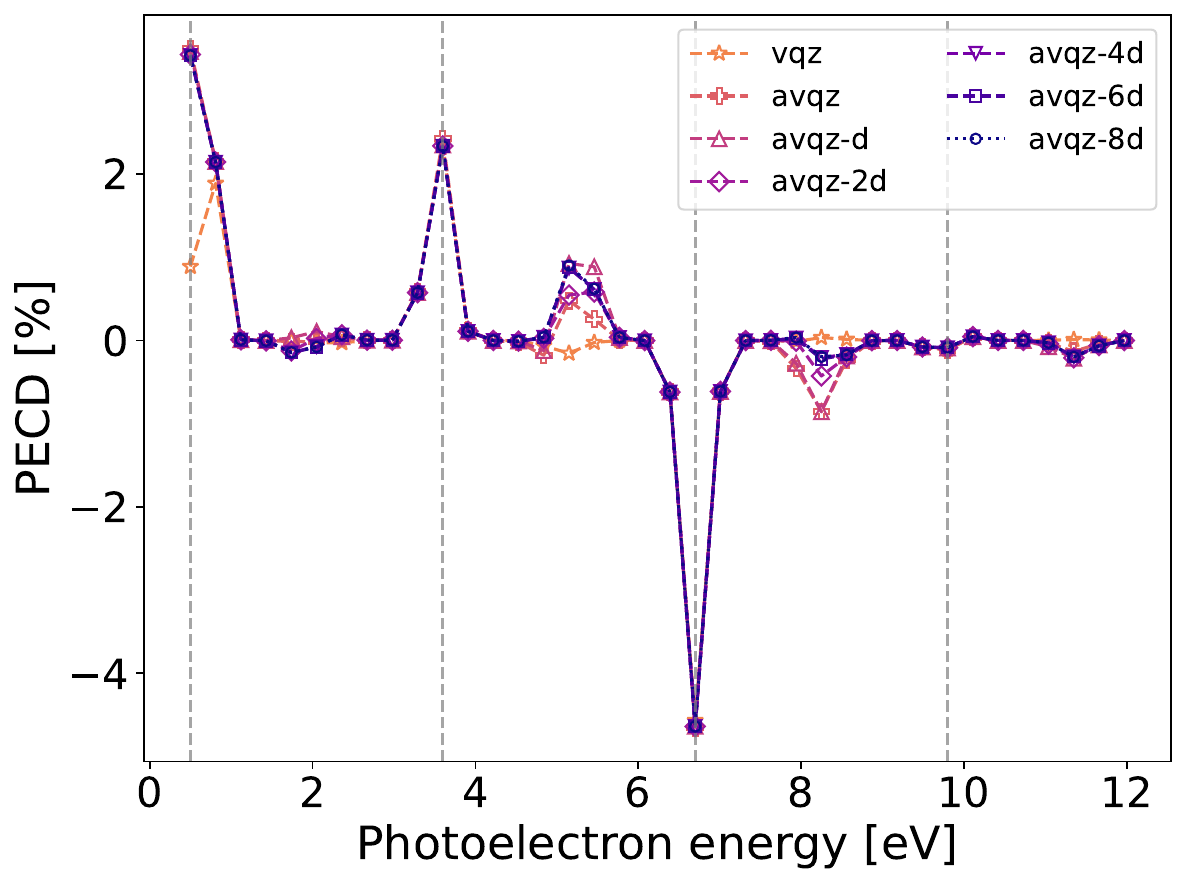}
	\caption{The PECD at $\theta = 0^{\circ}$ and $\tau_k =0$ is shown as a function of photoelectron energy. The vertical gray dashed lines indicate the harmonic positions. The calculation uses an XUV and IR pulses of 5$\,$fs FWHM. The figure demonstrates the convergence of PECD with respect to the diffuse functions. Compared to Fig.~\ref{PECDplot} in the manuscript, the convergence trend is consistent, but the PECD signal changes sign at both the harmonics and sidebands.
	}
	\label{pecd_theta0}
\end{figure}
\begin{figure}[tbp]
	\centering
	\includegraphics[width=\columnwidth]{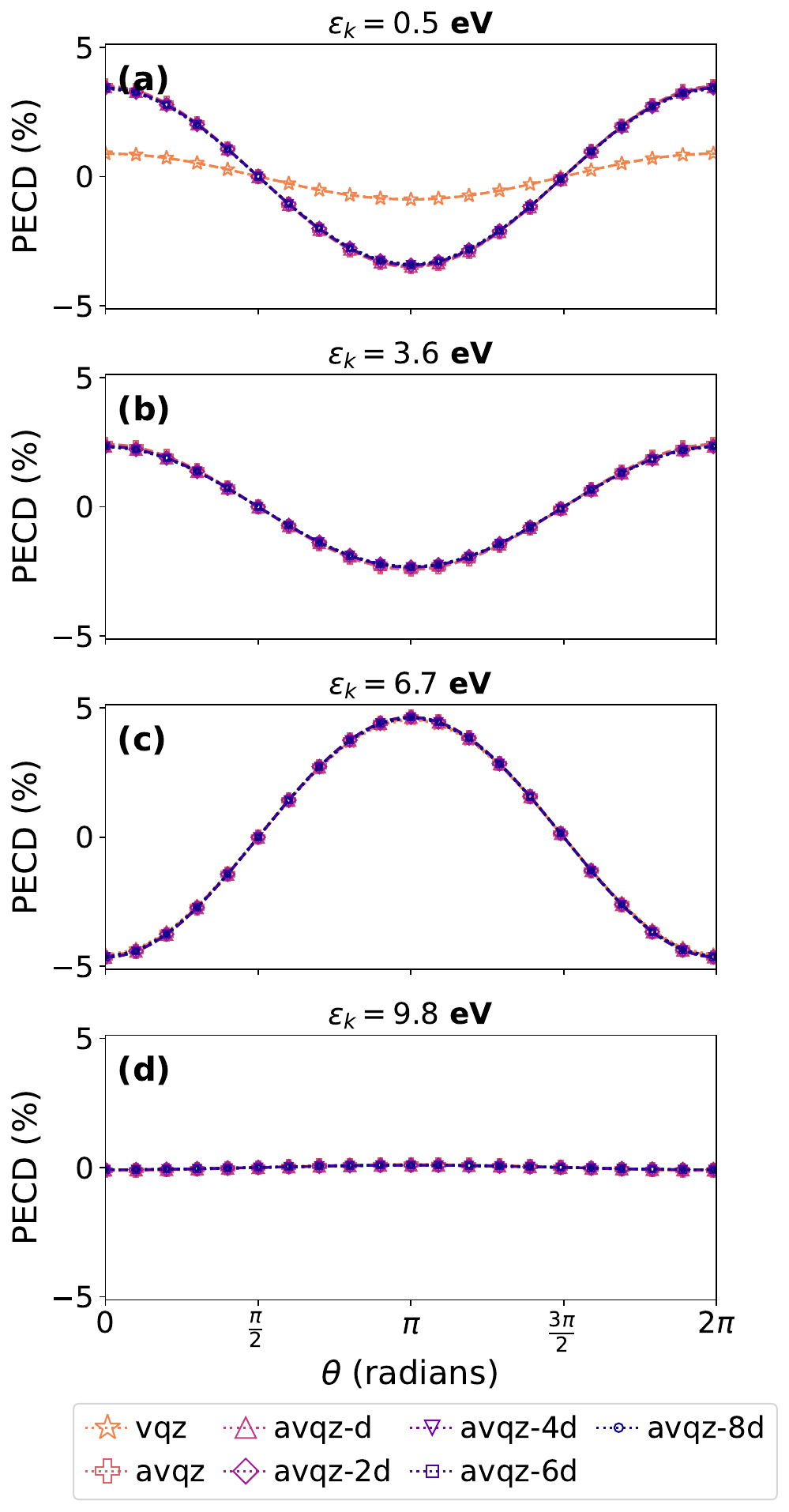} 
	
	\caption{PECD as a function of the photoelectron emission angle ($\theta$) at photoelectron energies corresponding to the harmonics: (a) H\,7, (b) H\,9, (c) H\,11, and (d) H\,13. The PECD at the harmonic energies is very well converged with respect to the diffuse functions across the full range of $\theta$.
	}
	\label{PECD_theta_H}
\end{figure}
\begin{figure}[tbp]
	\centering
	\includegraphics[width=\columnwidth]{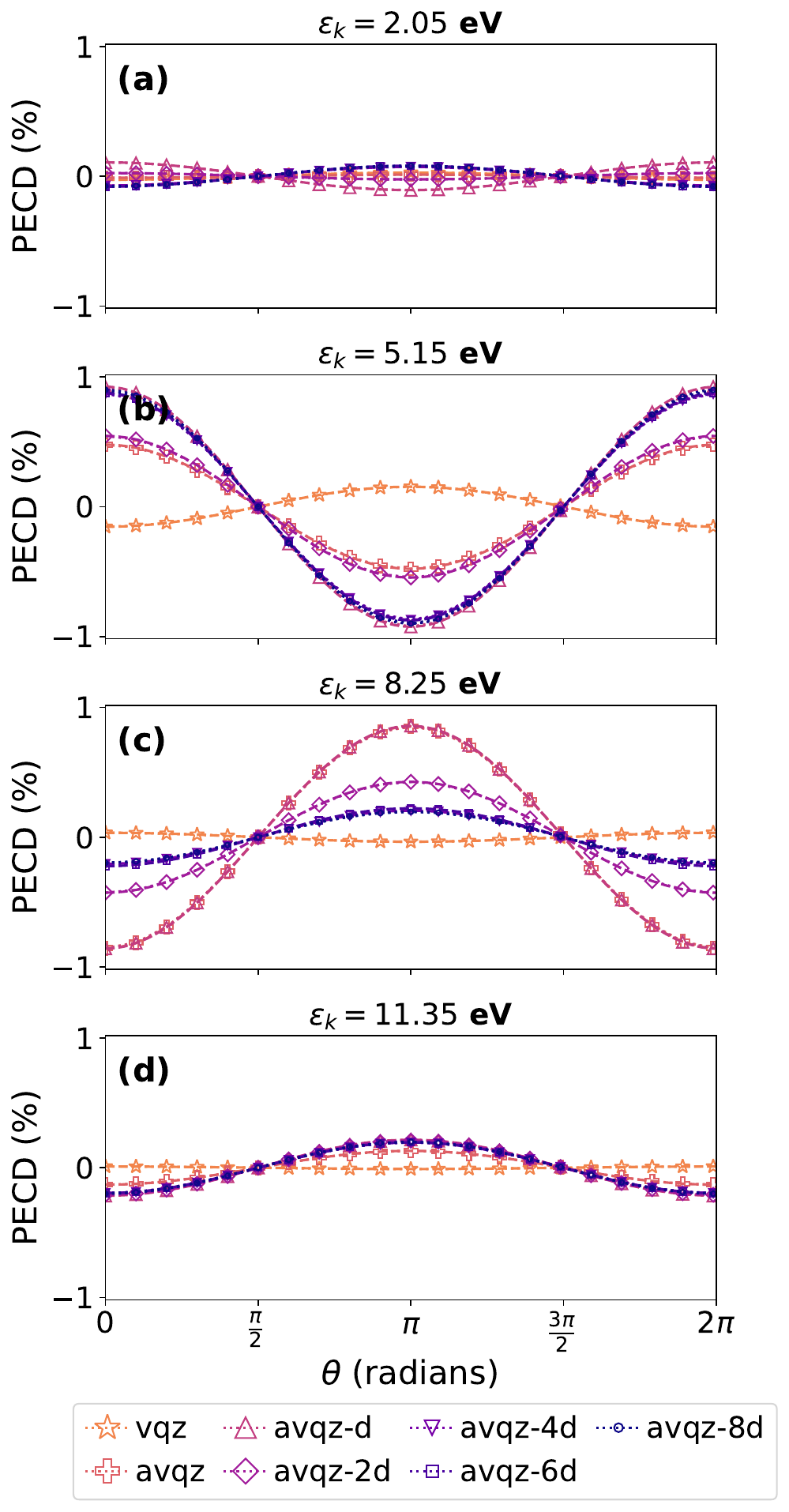} 
	\caption{PECD as a function of the photoelectron emission angle ($\theta$) at photoelectron energies corresponding to sidebands: (a) SB8, (b) SB10, (c) SB12, and (d) SB14. Compared to the harmonics, the PECD at sideband energies shows slower convergence with respect to the diffuse functions; however, it still converges across the full range of $\theta$.
	}
	\label{PECD_theta_SBs}
\end{figure}

Figures~\ref{PECD_theta_H} and \ref{PECD_theta_SBs} show the PECD as a function of the emission angle $\theta$ at the harmonic and sideband energies, respectively, complementing the fixed-angle results of Figs.~\ref{pecd_theta0} and \ref{PECDplot}. In both cases the angular dependence remains predominantly dipolar, with the asymmetry maximal along the light propagation axis. The convergence follows the trend observed at fixed angle, uniformly across the angular range: at the harmonics, a single set of diffuse functions essentially suffices, and only the unaugmented basis deviates at the lowest energy [Fig.~\ref{PECD_theta_H}(a)]; at the sidebands, the PECD curves differ visibly between the smaller basis sets and agree with the converged result only from avqz-4d onward (Fig.~\ref{PECD_theta_SBs}). At SB\,10, the unaugmented basis yields the opposite sign of the asymmetry over most of the angular range, underscoring that the diffuse functions are essential for even a qualitatively correct description of the two-photon PECD.

\begin{figure*}[tbp]
	\centering
	\includegraphics[width=\textwidth]{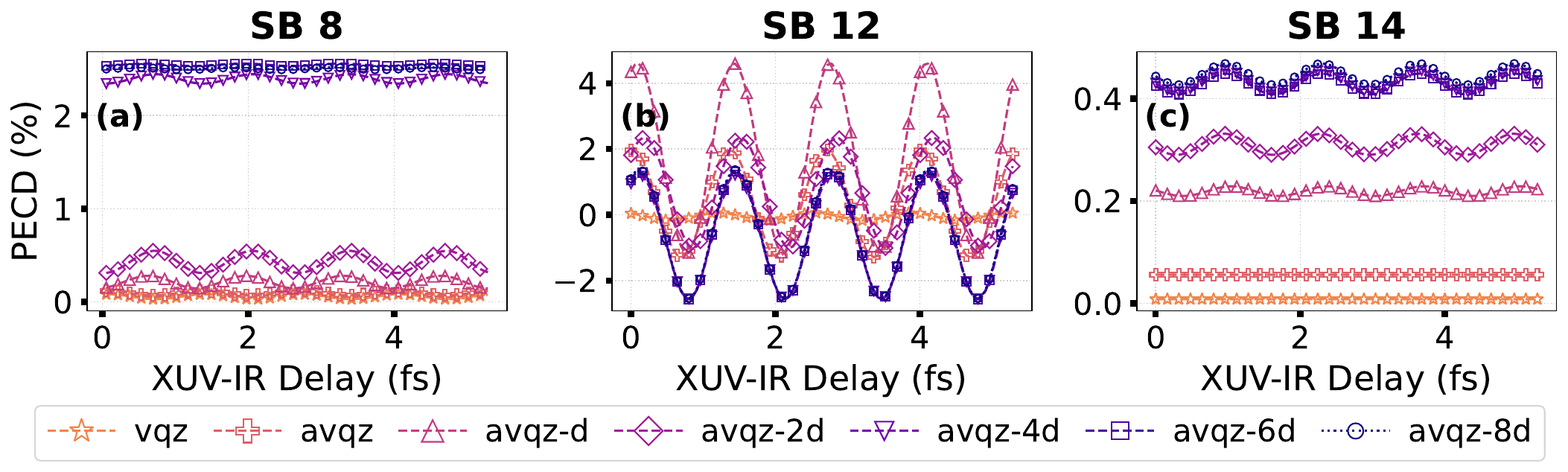} 
	
	\caption{Convergence of PECD as a function of the XUV–IR delay at photoelectron energies corresponding to sidebands (a) SB\,8, (b) SB\,12, and (c) SB\,14.
	}
	\label{delayed PECD SB}
\end{figure*}

Finally, Fig.~\ref{delayed PECD SB} shows the PECD convergence as a function of XUV–IR delay for SB 8, SB 12, and SB 14; the SB 10 results appear in the main text (Fig.~\ref{pecdtimedelay}). The convergence is consistent across all sidebands. The corresponding numerical values, listed in Table~\ref{delaypecd:table}, converge to at least two digits in all cases.

Taken together, these results, along with Figs.~\ref{PECDplot} and \ref{pecdtimedelay} of the main text, demonstrate that the PECD in our theoretical framework is well converged with respect to the inclusion of diffuse Gaussian functions. 

\begin{table}[tbp]
	\caption{Peak PECD (in \%) at sidebands SB\,8--SB\,14, computed with increasingly diffuse basis sets (vqz through avqz-8d). The data for SB\,8, SB\,12, and SB\,14 are shown in Fig.~\ref{delayed PECD SB}; those for SB\,10 are shown in Fig.~\ref{pecdtimedelay}. The peak PECD converges to at least two digits at all sidebands.}
	\label{delaypecd:table}
	\begin{ruledtabular}
		\begin{tabular}{lcccc}
			Basis set & SB\,8  & SB\,10 (Fig.~\ref{pecdtimedelay}) & SB\,12  & SB\,14 \\
			\colrule
			vqz     & 0.082   & 0.359  & 0.054  & 0.008 \\
			avqz    & 0.121   & 1.864  & 1.969  & 0.056 \\
			avqz-d  & 0.280   & 8.872  & 4.593  & 0.229 \\
			avqz-2d & 0.548   & 17.720 & 2.332  & 0.332 \\
			avqz-4d & 2.444   & 14.658 & 1.207  & 0.456 \\
			avqz-6d & 2.558   & 14.508 & 1.310  & 0.449 \\
			avqz-8d & 2.516   & 14.445 & 1.367  & 0.469 \\
		\end{tabular}
	\end{ruledtabular}
\end{table}

\section{Anisotropy parameters}
\label{app:betas}
\renewcommand{\thefigure}{B\arabic{figure}}
\setcounter{figure}{0}
\begin{table}[tbp]
	\caption{Nonzero anisotropy parameters $\beta_{LM}$ and their contributing
		ionization pathways [see Eq.~\protect\eqref{betasum}]. A check mark denotes
		presence, a cross absence. One-photon process is restricted to $L\le2$,
		so the parameters with $L=3,4$ arise solely from two-photon transitions. The total contribution to each parameter is the sum of
		the terms marked below.} 
	\label{beta_table}
	\begin{ruledtabular}
		\begin{tabular}{lcc}
			$\beta_{LM}$ & 1ph & 2ph \\
			\colrule
			$\beta_{00}$ & \cmark & \cmark \\
			$\beta_{10}$ & \cmark & \cmark \\
			$\beta_{20}$ & \cmark & \cmark \\
			$\beta_{22}^{*}=\beta_{2-2}$ & \xmark & \cmark \\
			$\beta_{30}$ & \xmark & \cmark \\
			$\beta_{32}^{*}=\beta_{3-2}$ & \xmark & \cmark \\
			$\beta_{40}$ & \xmark & \cmark \\
			$\beta_{42}^{*}=\beta_{4-2}$ & \xmark & \cmark \\
		\end{tabular}
	\end{ruledtabular}
\end{table}
\begin{table*}[tbp]
	\caption{Convergence of the anisotropy parameters $\beta_{30}$ and
		$\beta_{32}$ (real parts) with the number of added diffuse functions, for
		a 5\,fs (FWHM) XUV pulse and a 15\,fs (FWHM) IR pulse. Values are given at
		the sideband photoelectron energies, converged digits are highlighted in red.}
	\label{tab:beta3_conv}
	\begin{ruledtabular}
		\begin{tabular}{lcccccccc}
			& \multicolumn{4}{c}{$\beta_{30}$} & \multicolumn{4}{c}{$\beta_{32}=\beta_{3-2}$} \\
			\cline{2-5}\cline{6-9}
			Basis & 2.05 & 5.15 & 8.25 & 11.35 & 2.05 & 5.15 & 8.25 & 11.35 \\
			\colrule
			vqz     & -0.000141 & 0.000011 & -0.000122 & -0.000003 & 0.000129 & -0.000010 & \textcolor{red}{0.00}0111 & 0.000002 \\
			avqz    & -0.000319 & 0.000369 & -0.002433 & -0.000006 & 0.000291 & -0.000337 & \textcolor{red}{0.00}2221 & 0.000006 \\
			avqz-d  & 0.000273 & \textcolor{red}{-0.0}12551 & \textcolor{red}{-0.00}7654 & \textcolor{red}{0.00}0568 & -0.000249 & \textcolor{red}{0.0}11457 & \textcolor{red}{0.00}6987 & \textcolor{red}{-0.00}0519 \\
			avqz-2d & \textcolor{red}{-0.00}0634 & \textcolor{red}{-0.0}33317 & \textcolor{red}{-0.00}3579 & \textcolor{red}{0.00}0994 & \textcolor{red}{0.00}0578 & \textcolor{red}{0.0}30414 & \textcolor{red}{0.00}3268 & \textcolor{red}{-0.00}0907 \\
			avqz-4d & \textcolor{red}{-0.00}4469 & \textcolor{red}{-0.02}2458 & \textcolor{red}{-0.00}2051 & \textcolor{red}{0.00}1398 & \textcolor{red}{0.00}4079 & \textcolor{red}{0.0}20501 & \textcolor{red}{0.00}1872 & \textcolor{red}{-0.00}1276 \\
			avqz-6d & \textcolor{red}{-0.004}823 & \textcolor{red}{-0.02}1431 & \textcolor{red}{-0.002}200 & \textcolor{red}{0.001}384 & \textcolor{red}{0.004}402 & \textcolor{red}{0.0}19564 & \textcolor{red}{0.00}2010 & \textcolor{red}{-0.001}264 \\
			avqz-8d & \textcolor{red}{-0.004}717 & \textcolor{red}{-0.021}391 & \textcolor{red}{-0.002}422 & \textcolor{red}{0.001}162 & \textcolor{red}{0.004}306 & \textcolor{red}{0.0195}28 & \textcolor{red}{0.002}211 & \textcolor{red}{-0.001}061 \\
		\end{tabular}
	\end{ruledtabular}
\end{table*}

\begin{figure*}[!t]
	\centering
	\includegraphics[width=\textwidth]{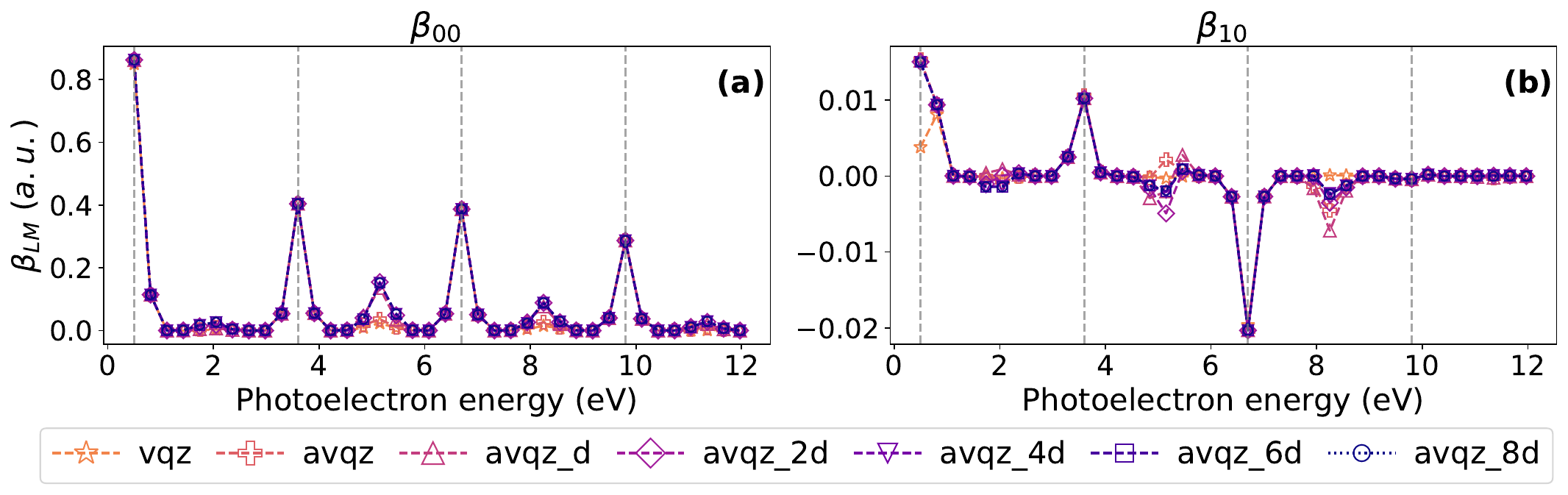}
	\caption{Anisotropy parameters $\beta_{00}$ and $\beta_{10}$ as functions of the photoelectron energy. Both contain contributions from one- and two-photon pathways: $\beta_{00}$ corresponds to the photoelectron spectrum (Eq.~\protect\eqref{PES}), while $\beta_{10}$ contributes to the PECD at harmonics (Eq.~\protect\eqref{pecd_H}).
	}
	\label{beta00L1}
\end{figure*}

\begin{figure*}[!t]
	\centering
	\includegraphics[width=\textwidth]{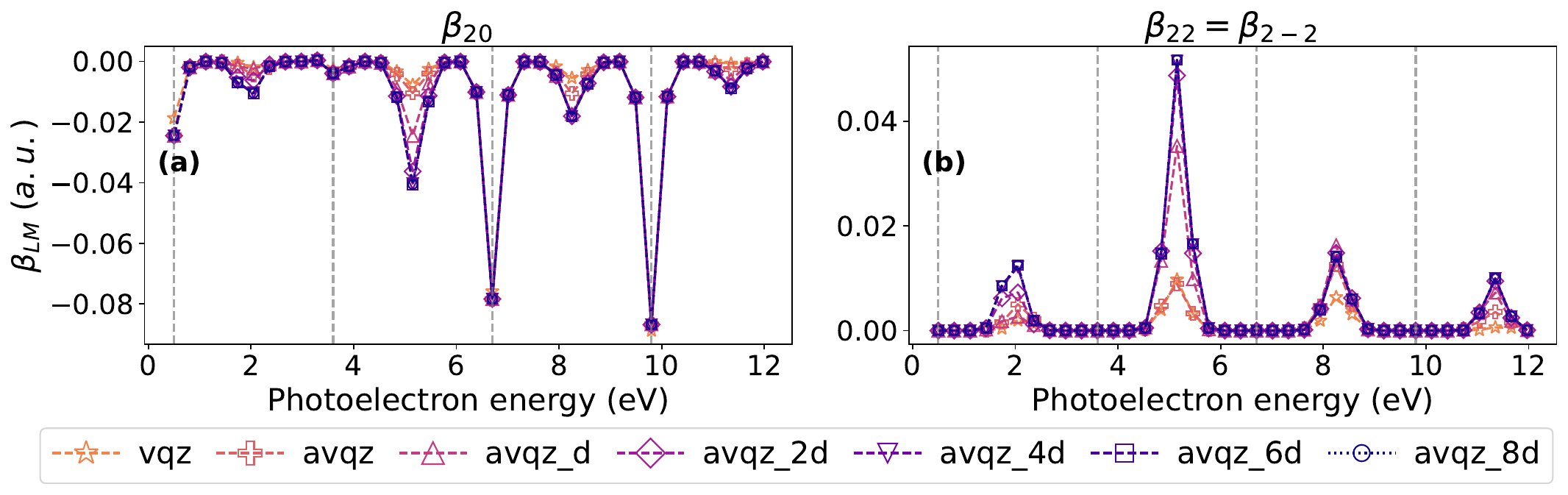} 
	\caption{Anisotropy parameters for $L=2$ as functions of the photoelectron energy: (a) $\beta_{20}$ $(M=0)$ and (b) $\beta_{22}$ $(M=2)$. The parameter $\beta_{20}$ contains contributions from both one- and two-photon pathways, while $\beta_{22}$ originate solely from two-photon pathways.}
	\label{betaL2}
\end{figure*}

\begin{figure*}[tb]
	\centering
	\includegraphics[width=\textwidth]{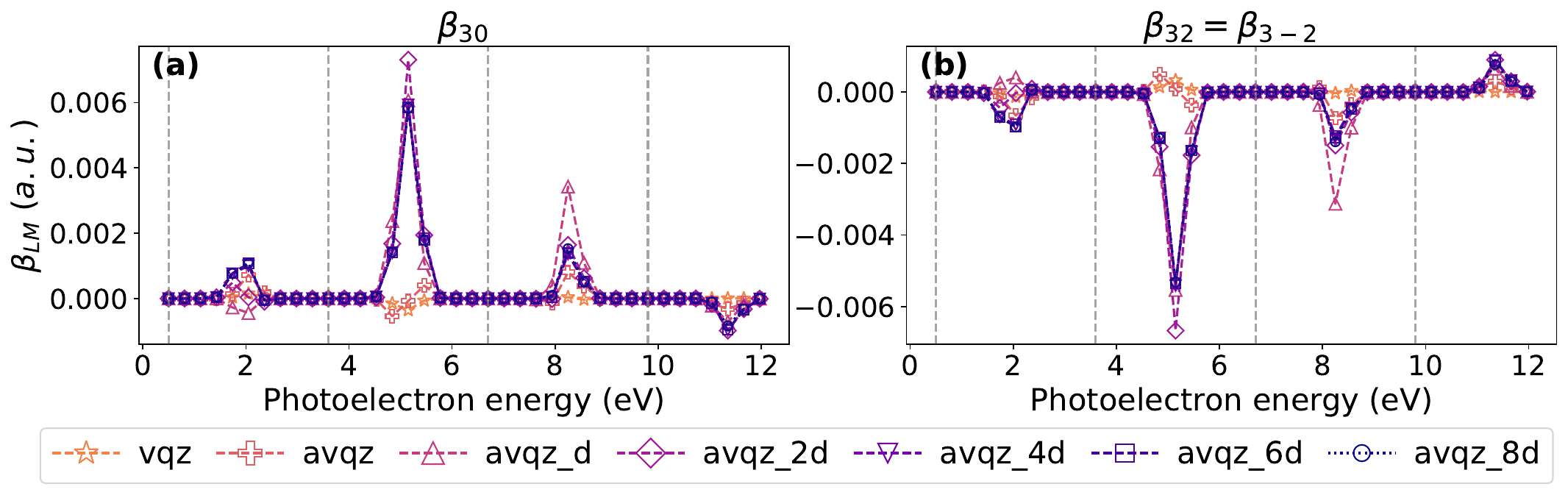}
	\caption{Anisotropy parameters for $L=3$ as functions of the photoelectron energy: (a) $\beta_{30}$ $(M=0)$ and (b) $\beta_{32}$ $(M=2)$. Both $\beta_{30}$ and $\beta_{32}$ originate exclusively from two-photon pathways. Here, $\beta_{32}$ satisfy the symmetry relation (see Eq.~\eqref{betasym}) such that $\beta_{32}=\beta_{3-2}$.}
	\label{betaL3}
\end{figure*}

\begin{figure*}[tb]
	\centering
	\includegraphics[width=\textwidth]{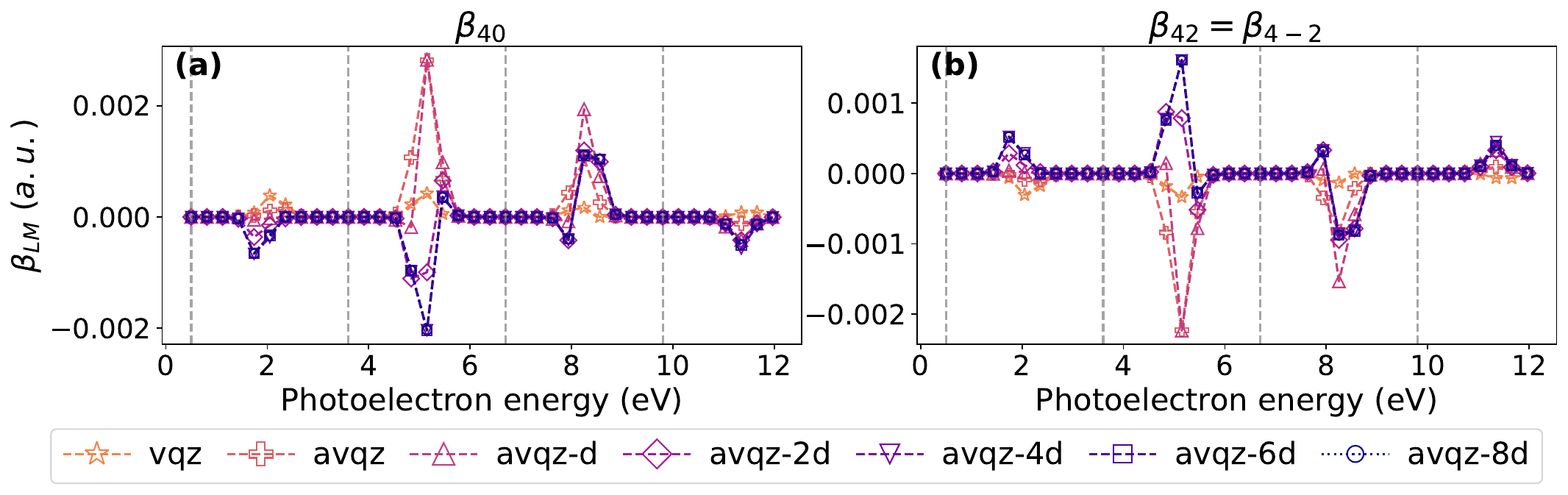} 
	\caption{Anisotropy parameters for $L=4$ as functions of the photoelectron energy: (a) $\beta_{40}$ $(M=0)$ and (b) $\beta_{42}$ $(M=2)$. Both $\beta_{40}$ and $\beta_{42}$ arise exclusively from two-photon photoionization pathways. Here, $\beta_{42}$ satisfy the symmetry relation (see Eq.~\eqref{betasym}) such that $\beta_{42}=\beta_{4-2}$.}
	\label{betaL4}
\end{figure*}

\begin{table*}[tbh]
	\caption{Convergence of the anisotropy parameters $\beta_{40}$ and
		$\beta_{42}$ (real parts) with the number of added diffuse functions at sidebands, for
		a 5\,fs (FWHM) XUV pulse and a 15\,fs (FWHM) IR pulse, converged digits are highlighted in red.}
	\label{tab:beta4_conv}
	\begin{ruledtabular}
		\begin{tabular}{lcccccccc}
			& \multicolumn{4}{c}{$\beta_{40}$} & \multicolumn{4}{c}{$\beta_{42}=\beta_{4-2}$} \\
			\cline{2-5}\cline{6-9}
			Basis & 2.05 & 5.15 & 8.25 & 11.35 & 2.05 & 5.15 & 8.25 & 11.35 \\
			\colrule
			vqz     & 0.000296 & -0.000317 & -0.000362 & 0.000063 & -0.000234 & 0.000251 & 0.000286 & -0.000050 \\
			avqz    & 0.000029 & 0.013663 & 0.005616 & 0.000072 & -0.000023 & -0.010801 & -0.004440 & -0.000057 \\
			avqz-d  & \textcolor{red}{0.00}0451 & 0.010619 & \textcolor{red}{0.0}16568 & \textcolor{red}{-0.00}0268 & \textcolor{red}{-0.00}0356 & -0.008395 & \textcolor{red}{-0.0}13098 & \textcolor{red}{0.00}0212 \\
			avqz-2d & \textcolor{red}{0.00}1484 & 0.000944 & \textcolor{red}{0.01}0001 & \textcolor{red}{-0.00}1263 & \textcolor{red}{-0.00}1173 & -0.000747 & \textcolor{red}{-0.00}7907 & \textcolor{red}{0.00}0200 \\
			avqz-4d & \textcolor{red}{0.000}651 & \textcolor{red}{-0.00}4847 & \textcolor{red}{0.01}2047 & \textcolor{red}{-0.00}2658 & \textcolor{red}{-0.00}0515 & \textcolor{red}{0.00}3832 & \textcolor{red}{-0.00}9524 & \textcolor{red}{0.00}2101 \\
			avqz-6d & \textcolor{red}{0.0007}37 & \textcolor{red}{-0.004}928 & \textcolor{red}{0.0122}36 & \textcolor{red}{-0.002}541 & \textcolor{red}{-0.0005}83 & \textcolor{red}{0.003}896 & \textcolor{red}{-0.009}673 & \textcolor{red}{0.002}009 \\
			avqz-8d & \textcolor{red}{0.0007}18 & \textcolor{red}{-0.004}771 & \textcolor{red}{0.0122}84 & \textcolor{red}{-0.0025}30 & \textcolor{red}{-0.0005}68 & \textcolor{red}{0.003}772 & \textcolor{red}{-0.009}711 & \textcolor{red}{0.00200}1 \\
		\end{tabular}
	\end{ruledtabular}
\end{table*}
Table~\ref{beta_table} lists all nonzero anisotropy parameters for circularly polarized XUV and linearly polarized IR fields, together with their contributions from the one- and two-photon pathways. Here, we examine the convergence of the real parts of the anisotropy parameters of
Table~\ref{beta_table} with respect to the inclusion of diffuse functions for
a shorter IR pulse. In the main text, the convergence of all nonzero
parameters is reported for a 5\,fs (FWHM) XUV pulse and a 15\,fs (FWHM) IR
pulse; here, we use the same XUV pulse with a 5\,fs (FWHM) IR pulse to assess
the effect of the pulse duration. Figures~\ref{beta00L1}--\ref{betaL4} show
the anisotropy parameters for $L=0$--$4$ versus the photoelectron energy. The
convergence behavior closely parallels that of the main text: the parameters
containing a one-photon component converge rapidly at the harmonics, to four
to six decimal places, while the convergence at the sidebands is slower and
more uneven. At the sidebands, $\beta_{00}$ converges to five decimal places
at SB\,8, three at SB\,10 and SB\,12, and four at SB\,14; $\beta_{10}$ to five
decimal places at SB\,8 and SB\,12, and three at SB\,10 and SB\,14;
$\beta_{20}$ to four decimal places at SB\,8, SB\,12, and SB\,14, and five at
SB\,10; and $\beta_{22}$ to three decimal places at SB\,8--SB\,12, and two at
SB\,14. The higher-order parameters behave similarly: $\beta_{30}$ converges
to five decimal places at SB\,8, four at SB\,10, and three at SB\,12 and
SB\,14; $\beta_{32}$ to four decimal places at SB\,8 and three at
SB\,10--SB\,14; $\beta_{40}$ to four decimal places at SB\,8--SB\,12 and five
at SB\,14; and $\beta_{42}$ to three decimal places at SB\,8 and SB\,14, four
at SB\,10, and five at SB\,12. The convergence is thus comparable to that
obtained with the 15\,fs IR pulse, both in the number of converged decimal
places and in the harmonic--sideband contrast, confirming that it is
essentially independent of the IR pulse duration.\\

\FloatBarrier
\bibliography{BasisSet}

@article{pazourek2015RMP,
	title = {Attosecond chronoscopy of photoemission},
	author = {Pazourek, Renate and Nagele, Stefan and Burgd\"orfer, Joachim},
	journal = {Rev. Mod. Phys.},
	volume = {87},
	issue = {3},
	pages = {765--802},
	numpages = {38},
	year = {2015},
	month = {Aug},
	publisher = {American Physical Society},
	doi = {10.1103/RevModPhys.87.765},
	url = {https://link.aps.org/doi/10.1103/RevModPhys.87.765}
}

@article{nisoli2017ChemRev,
	author = {Nisoli, Mauro and Decleva, Piero and Calegari, Francesca and Palacios, Alicia and Martín, Fernando},
	title = {Attosecond Electron Dynamics in Molecules},
	journal = {Chemical Reviews},
	volume = {117},
	number = {16},
	pages = {10760-10825},
	year = {2017},
	doi = {10.1021/acs.chemrev.6b00453},
	note ={PMID: 28488433},
	URL = {https://doi.org/10.1021/acs.chemrev.6b00453},
	eprint = {https://doi.org/10.1021/acs.chemrev.6b00453	}
}

@article{schultze2010Science,
	author = {M. Schultze  and M. Fieß  and N. Karpowicz  and J. Gagnon  and M. Korbman  and M. Hofstetter  and S. Neppl  and A. L. Cavalieri  and Y. Komninos  and Th. Mercouris  and C. A. Nicolaides  and R. Pazourek  and S. Nagele  and J. Feist  and J. Burgdörfer  and A. M. Azzeer  and R. Ernstorfer  and R. Kienberger  and U. Kleineberg  and E. Goulielmakis  and F. Krausz  and V. S. Yakovlev },
	title = {Delay in Photoemission},
	journal = {Science},
	volume = {328},
	number = {5986},
	pages = {1658-1662},
	year = {2010},
	doi = {10.1126/science.1189401},
	URL = {https://www.science.org/doi/abs/10.1126/science.1189401},
	eprint = {https://www.science.org/doi/pdf/10.1126/science.1189401}
}

@article{klunder2011PRL,
	title = {Probing Single-Photon Ionization on the Attosecond Time Scale},
	author = {Kl\"under, K. and Dahlstr\"om, J. M. and Gisselbrecht, M. and Fordell, T. and Swoboda, M. and Gu\'enot, D. and Johnsson, P. and Caillat, J. and Mauritsson, J. and Maquet, A. and Ta\"{\i}eb, R. and L'Huillier, A.},
	journal = {Phys. Rev. Lett.},
	volume = {106},
	issue = {14},
	pages = {143002},
	numpages = {4},
	year = {2011},
	month = {Apr},
	publisher = {American Physical Society},
	doi = {10.1103/PhysRevLett.106.143002},
	url = {https://link.aps.org/doi/10.1103/PhysRevLett.106.143002}
}

@article{isinger2017Science,
	author = {M. Isinger  and R. J. Squibb  and D. Busto  and S. Zhong  and A. Harth  and D. Kroon  and S. Nandi  and C. L. Arnold  and M. Miranda  and J. M. Dahlström  and E. Lindroth  and R. Feifel  and M. Gisselbrecht  and A. L’Huillier },
	title = {Photoionization in the time and frequency domain},
	journal = {Science},
	volume = {358},
	number = {6365},
	pages = {893-896},
	year = {2017},
	doi = {10.1126/science.aao7043},
	URL = {https://www.science.org/doi/abs/10.1126/science.aao7043},
	eprint = {https://www.science.org/doi/pdf/10.1126/science.aao7043}

}

@article{goetz2016PRA,
	title = {Maximizing hole coherence in ultrafast photoionization of argon with an optimization by sequential parametrization update},
	author = {Goetz, R. Esteban and Merkel, Maximilian and Karamatskou, Antonia and Santra, Robin and Koch, Christiane P.},
	journal = {Phys. Rev. A},
	volume = {94},
	issue = {2},
	pages = {023420},
	numpages = {12},
	year = {2016},
	month = {Aug},
	publisher = {American Physical Society},
	doi = {10.1103/PhysRevA.94.023420},
	url = {https://link.aps.org/doi/10.1103/PhysRevA.94.023420}
}

@article{douguet2018PRA,
	title = { of the complex Kohn variational method to attosecond spectroscopy},
	author = {Douguet, N. and Schneider, B. I. and Argenti, L.},
	journal = {Phys. Rev. A},
	volume = {98},
	issue = {2},
	pages = {023403},
	numpages = {12},
	year = {2018},
	month = {Aug},
	publisher = {American Physical Society},
	doi = {10.1103/PhysRevA.98.023403},
	url = {https://link.aps.org/doi/10.1103/PhysRevA.98.023403}
}

@article{stener2004JCP,
	author  = {Stener, Mauro and Fronzoni, Giovanni and Di Tommaso, Daniela and Decleva, Piero},
	title   = {Density Functional Study on the Circular Dichroism of Photoelectron Angular Distribution from Chiral Derivatives of Oxirane},
	journal = {The Journal of Chemical Physics},
	year    = {2004},
	volume   = {120},
	number   = {7},
	pages    = {3284--3296},
	doi      = {10.1063/1.1640617}
}

@article{demekhin2015JCP,
	author  = {Artemyev, Anton N. and M{\"u}ller, Anne D. and Hochstuhl, David and Demekhin, Philipp V.},
	title   = {Photoelectron Circular Dichroism in the Multiphoton Ionization by Short Laser Pulses. I. Propagation of Single-Active-Electron Wave Packets in Chiral Pseudo-Potentials},
	journal = {The Journal of Chemical Physics},
	year    = {2015},
	volume  = {142},
	number  = {24},
	pages   = {244105},
	doi     = {10.1063/1.4922690}
}

@book{mcnaught1997GoldBook,
	author    = {McNaught, A. D. and Wilkinson, A.},
	title     = {Compendium of Chemical Terminology: IUPAC Recommendations},
	publisher = {Blackwell Science},
	address   = {Oxford},
	year      = {1997},
	note      = {The "Gold Book"},
}

@article{ma2020CSBB,
	author  = {Ma, Yingyi and Shi, Lei and Yue, Hongyan and Gao, Xin},
	title   = {Recognition at chiral interfaces: From molecules to cells},
	journal = {Colloids and Surfaces B: Biointerfaces},
	volume  = {195},
	pages   = {111268},
	year    = {2020},
	doi     = {10.1016/j.colsurfb.2020.111268},
	issn    = {0927-7765},
}

@article{liu2015ChemRev,
	author  = {Liu, Minghua and Zhang, Li and Wang, Tianyu},
	title   = {Supramolecular Chirality in Self-Assembled Systems},
	journal = {Chemical Reviews},
	volume  = {115},
	number  = {15},
	pages   = {7304--7397},
	year    = {2015},
	doi     = {10.1021/cr500671p},
	note    = {PMID: 26189453},
	url     = {https://doi.org/10.1021/cr500671p}
}

@article{fiechter2023StructDyn,
	author  = {Fiechter, Marit R. and Svoboda, Vít and Wörner, Hans Jakob},
	title   = {Theoretical study of time-resolved photoelectron circular dichroism in the photodissociation of a chiral molecule},
	journal = {Structural Dynamics},
	volume  = {10},
	number  = {6},
	pages   = {064103},
	year    = {2023},
	month   = {12},
	doi     = {10.1063/4.0000213},
	issn    = {2329-7778}
}

@article{nahon2015JESRP,
	author  = {Nahon, Laurent and Garcia, Gustavo A. and Powis, Ivan},
	title   = {Valence shell one-photon photoelectron circular dichroism in chiral systems},
	journal = {Journal of Electron Spectroscopy and Related Phenomena},
	volume  = {204},
	pages   = {322--334},
	year    = {2015},
	doi     = {10.1016/j.elspec.2015.04.008},
	issn    = {0368-2048}
}

@article{ritchie1976PhysRevA,
	author  = {Ritchie, Burke},
	title   = {Theory of the angular distribution of photoelectrons ejected from optically active molecules and molecular negative ions},
	journal = {Physical Review A},
	volume  = {13},
	number  = {4},
	pages   = {1411--1415},
	year    = {1976},
	month   = {Apr},
	doi     = {10.1103/PhysRevA.13.1411}
}

@article{powis2000JCP,
	author  = {Powis, Ivan},
	title   = {Photoelectron circular dichroism of the randomly oriented chiral molecules glyceraldehyde and lactic acid},
	journal = {The Journal of Chemical Physics},
	volume  = {112},
	number  = {1},
	pages   = {301--310},
	year    = {2000},
	month   = {Jan},
	doi     = {10.1063/1.480581}
}

@article{boewering2001PRL,
	author  = {Böwering, N. and Lischke, T. and Schmidtke, B. and Müller, N. and Khalil, T. and Heinzmann, U.},
	title   = {Asymmetry in Photoelectron Emission from Chiral Molecules Induced by Circularly Polarized Light},
	journal = {Physical Review Letters},
	volume  = {86},
	number  = {7},
	pages   = {1187--1190},
	year    = {2001},
	month   = {Feb},
	doi     = {10.1103/PhysRevLett.86.1187}
}

@article{waters2022CPC,
	author  = {Waters, Max D. J. and Ladda, Nicolas and Senftleben, Arne and Svoboda, Vít and Belozertsev, Mikhail and Baumert, Thomas and Wörner, Hans Jakob},
	title   = {Ground-State Photoelectron Circular Dichroism of Methyl p-Tolyl Sulfoxide by Single-Photon Ionisation from a Table-Top Source},
	journal = {ChemPhysChem},
	volume  = {23},
	number  = {24},
	pages   = {e202200575},
	year    = {2022},
	doi     = {10.1002/cphc.202200575}
}

@article{lux2012ACIE,
	author  = {Lux, Christian and Wollenhaupt, Matthias and Bolze, Tom and Liang, Qingqing and Köhler, Jens and Sarpe, Cristian and Baumert, Thomas},
	title   = {Circular Dichroism in the Photoelectron Angular Distributions of Camphor and Fenchone from Multiphoton Ionization with Femtosecond Laser Pulses},
	journal = {Angewandte Chemie International Edition},
	volume  = {51},
	number  = {20},
	pages   = {5001--5005},
	year    = {2012},
	doi     = {10.1002/anie.201109035}
}

@article{lux2015CPC,
	author  = {Lux, Christian and Wollenhaupt, Matthias and Sarpe, Cristian and Baumert, Thomas},
	title   = {Photoelectron Circular Dichroism of Bicyclic Ketones from Multiphoton Ionization with Femtosecond Laser Pulses},
	journal = {ChemPhysChem},
	volume  = {16},
	number  = {1},
	pages   = {115--137},
	year    = {2015},
	doi     = {10.1002/cphc.201402643}
}

@article{lux2016JPhysB,
	author  = {Lux, Christian and Senftleben, Arne and Sarpe, Cristian and Wollenhaupt, Matthias and Baumert, Thomas},
	title   = {Photoelectron circular dichroism observed in the above-threshold ionization signal from chiral molecules with femtosecond laser pulses},
	journal = {Journal of Physics B: Atomic, Molecular and Optical Physics},
	volume  = {49},
	number  = {2},
	pages   = {02LT01},
	year    = {2016},
	doi     = {10.1088/0953-4075/49/2/02LT01}
}

@article{kastner2016CPC,
	author = {Kastner, Alexander and Lux, Christian and Ring, Tom and Züllighoven, Stefanie and Sarpe, Cristian and Senftleben, Arne and Baumert, Thomas},
	title = {Enantiomeric Excess Sensitivity to Below One Percent by Using Femtosecond Photoelectron Circular Dichroism},
	journal = {ChemPhysChem},
	volume = {17},
	number = {8},
	pages = {1119-1122},
	doi = {10.1002/cphc.201501067},
	year = {2016}
}

@article{kastner2017JCP,
	author = {Kastner, Alexander and Ring, Tom and Krüger, Bastian C. and Park, G. Barratt and Schäfer, Tim and Senftleben, Arne and Baumert, Thomas},
	title = {Intermediate state dependence of the photoelectron circular dichroism of fenchone observed via femtosecond resonance-enhanced multi-photon ionization},
	journal = {The Journal of Chemical Physics},
	volume = {147},
	number = {1},
	pages = {013926},
	year = {2017},
	month = {05},
	issn = {0021-9606},
	doi = {10.1063/1.4982614},
}

@article{lehmann2013JCP,
	author = {Lehmann, C. Stefan and Ram, N. Bhargava and Powis, Ivan and Janssen, Maurice H. M.},
	title = {Imaging photoelectron circular dichroism of chiral molecules by femtosecond multiphoton coincidence detection},
	journal = {The Journal of Chemical Physics},
	volume = {139},
	number = {23},
	pages = {234307},
	year = {2013},
	month = {12},
	issn = {0021-9606},
	doi = {10.1063/1.4844295},
	url = {https://doi.org/10.1063/1.4844295}
}

@article{janssen2014PCCP,
	author ={Janssen, Maurice H. M. and Powis, Ivan},
	title  ={Detecting chirality in molecules by imaging photoelectron circular dichroism},
	journal  ={Phys. Chem. Chem. Phys.},
	year  ={2014},
	volume  ={16},
	issue  ={3},
	pages  ={856-871},
	publisher  ={The Royal Society of Chemistry},
	doi  ={10.1039/C3CP53741B},
	url  ={http://dx.doi.org/10.1039/C3CP53741B}
}

@article{fanood2015NatCom,
	author       = {Mohammad M. Rafiee Fanood and N. Bhargava Ram and C. Stefan Lehmann and Ivan Powis and Maurice H. M. Janssen},
	title        = {Enantiomer-specific analysis of multi-component mixtures by correlated electron imaging–ion mass spectrometry},
	journal      = {Nature Communications},
	volume       = {6},
	number       = {Article no. 7511},
	year         = {2015},
	doi          = {10.1038/ncomms8511},
	url          = {https://doi.org/10.1038/ncomms8511}
}

@article{goetz2017JCP,
	author  = {Goetz, R. E. and Isaev, T. A. and Nikoobakht, B. and Berger, R. and Koch, C. P.},
	title   = {Theoretical description of circular dichroism in photoelectron angular distributions of randomly oriented chiral molecules after multi-photon photoionization},
	journal = {The Journal of Chemical Physics},
	volume  = {146},
	number  = {2},
	pages   = {024306},
	year    = {2017},
	doi     = {10.1063/1.4973456}
}

@article{hergenhahn2004JCP,
	author  = {Hergenhahn, Uwe and Rennie, Emma E. and Kugeler, Oliver and Marburger, Simon and Lischke, Toralf and Powis, Ivan and Garcia, Gustavo},
	title   = {Photoelectron circular dichroism in core level ionization of randomly oriented pure enantiomers of the chiral molecule camphor},
	journal = {The Journal of Chemical Physics},
	volume  = {120},
	number  = {10},
	pages   = {4553--4556},
	year    = {2004},
	month   = {Mar},
	doi     = {10.1063/1.1651474}
}

@article{beaulieu2017Science,
	author  = {Beaulieu, S. and Comby, A. and Clergerie, A. and Caillat, J. and Descamps, D. and Dudovich, N. and Fabre, B. and Géneaux, R. and Légaré, F. and Petit, S. and Pons, B. and Porat, G. and Ruchon, T. and Taïeb, R. and Blanchet, V. and Mairesse, Y.},
	title   = {Attosecond-resolved photoionization of chiral molecules},
	journal = {Science},
	volume  = {358},
	number  = {6368},
	pages   = {1288--1294},
	year    = {2017},
	doi 	= {10.1126/science.aao5624},
	URL 	= {https://www.science.org/doi/abs/10.1126/science.aao5624},
}

@article{goetz2019JCP,
	author = {Goetz, R. Esteban and Koch, Christiane P. and Greenman, Loren},
	title = {Perfect control of photoelectron anisotropy for randomly oriented ensembles of molecules by XUV REMPI and polarization shaping},
	journal = {The Journal of Chemical Physics},
	volume = {151},
	number = {7},
	pages = {074106},
	year = {2019},
	month = {08},
	issn = {0021-9606},
	doi = {10.1063/1.5111362},
	url = {https://doi.org/10.1063/1.5111362},
}

@article{goetz2019PRL,
	author    = {Goetz, R. Esteban and Koch, Christiane P. and Greenman, Loren},
	title     = {Quantum control of photoelectron circular dichroism},
	journal   = {Physical Review Letters},
	year      = {2019},
	volume    = {122},
	number    = {1},
	pages     = {013204},
	publisher = {American Physical Society},
	doi       = {10.1103/PhysRevLett.122.013204},
	url       = {https://link.aps.org/doi/10.1103/PhysRevLett.122.013204}
}

@article{tikhonov2022SciAdv,
	author = {Denis S. Tikhonov  and Alexander Blech  and Monika Leibscher  and Loren Greenman  and Melanie Schnell  and Christiane P. Koch },
	title = {Pump-probe spectroscopy of chiral vibrational dynamics},
	journal = {Science Advances},
	volume = {8},
	number = {49},
	pages = {eade0311},
	year = {2022},
	doi = {10.1126/sciadv.ade0311}
}

@article{hanus2023RSC,
	author ={Hanus, Václav and Kangaparambil, Sarayoo and Richter, Martin and Haßfurth, Lukas and Dorner-Kirchner, Martin and Paulus, Gerhard G. and Xie, Xinhua and Baltuška, Andrius and Gräfe, Stefanie and Zeiler, Markus},
	title  		={Carrier envelope phase sensitivity of photoelectron circular dichroism},
	journal 	={Phys. Chem. Chem. Phys.},
	year  		={2023},
	volume  	={25},
	issue  		={6},
	pages  		={4656-4666},
	publisher	={The Royal Society of Chemistry},
	doi  		={10.1039/D2CP03077B}
}

@article{goetz2025PRR,
	title = {Continuum-electron interferometry for enhancement of photoelectron circular dichroism and measurement of bound, free, and mixed contributions to chiral response},
	author = {Goetz, R. Esteban and Blech, Alexander and Allison, Corbin and Koch, Christiane P. and Greenman, Loren},
	journal = {Phys. Rev. Res.},
	volume = {7},
	issue = {3},
	pages = {L032036},
	year = {2025},
	month = {Aug},
	publisher = {American Physical Society},
	doi = {10.1103/8f7h-6nfc},
	url = {https://link.aps.org/doi/10.1103/8f7h-6nfc}
}

@phdthesis{Blech2025,
	author = {Blech, Alexander},
	year = {2025},
	title = {Detection and control of chiral molecules},
	type = {Dissertation},
	url = {https://refubium.fu-berlin.de/handle/fub188/52078}
}

@article{Lucchese1982PRA,
	title = {Studies of differential and total photoionization cross sections of molecular nitrogen},
	author = {Lucchese, Robert R. and Raseev, Georges and McKoy, Vincent},
	journal = {Phys. Rev. A},
	volume = {25},
	issue = {5},
	pages = {2572--2587},
	year = {1982},
	month = {May},
	publisher = {American Physical Society},
	doi = {10.1103/PhysRevA.25.2572},
	url = {https://link.aps.org/doi/10.1103/PhysRevA.25.2572}
}

@misc{werner2012molpro,
	author    = {Werner, Hans-Joachim and Knowles, Peter J. and Knizia, Gerald and Manby, Frederick R. and Sch{\"u}tz, Martin and others},
	title     = {MOLPRO, version 2012.1, a package of ab initio programs},
	howpublished = {\url{http://www.molpro.net}},
	year      = {2012}
}

@article{werner2012WIRCMS,
	author    = {Werner, Hans-Joachim and Knowles, Peter J. and Knizia, Gerald and Manby, Frederick R. and Sch{\"u}tz, Martin},
	title     = {Molpro: A general-purpose quantum chemistry program package},
	journal   = {WIREs Computational Molecular Science},
	volume    = {2},
	number    = {2},
	pages     = {242--253},
	year      = {2012},
	doi       = {10.1002/wcms.82},
	url       = {https://doi.org/10.1002/wcms.82}
}

@article{muller2002APB,
	author  = {Muller, H. G.},
	title   = {Reconstruction of attosecond harmonic beating by interference of two-photon transitions},
	journal = {Applied Physics B},
	volume  = {74},
	pages   = {s17--s21},
	year    = {2002},
	month   = {Jun},
	doi     = {10.1007/s00340-002-0894-8}
}

@article{paul2001Science,
	author  = {Paul, P. M. and Toma, E. S. and Breger, P. and Mullot, G. and Augé, F. and Balcou, Ph. and Muller, H. G. and Agostini, P.},
	title   = {Observation of a Train of Attosecond Pulses from High Harmonic Generation},
	journal = {Science},
	volume  = {292},
	number  = {5522},
	pages   = {1689--1692},
	year    = {2001},
	doi     = {10.1126/science.1059413}
}

@article{McCurdy2001PRA63,
	title = {Practical calculation of amplitudes for electron-impact ionization},
	author = {McCurdy, C. William and Horner, Daniel A. and Rescigno, Thomas N.},
	journal = {Phys. Rev. A},
	volume = {63},
	issue = {2},
	pages = {022711},
	year = {2001},
	month = {Jan},
	publisher = {American Physical Society},
	doi = {10.1103/PhysRevA.63.022711},
	url = {https://link.aps.org/doi/10.1103/PhysRevA.63.022711}
}

@article{McCurdy2001PRA64,
	title = {Accurate amplitudes for electron-impact ionization},
	author = {Baertschy, M. and Rescigno, T. N. and McCurdy, C. W.},
	journal = {Phys. Rev. A},
	volume = {64},
	issue = {2},
	pages = {022709},
	numpages = {11},
	year = {2001},
	month = {Jul},
	publisher = {American Physical Society},
	doi = {10.1103/PhysRevA.64.022709},
	url = {https://link.aps.org/doi/10.1103/PhysRevA.64.022709}
}

@article{dahlstrom2013ChemPhys,
	title = {Theory of attosecond delays in laser-assisted photoionization},
	journal = {Chemical Physics},
	volume = {414},
	pages = {53--64},
	year = {2013},
	note = {Attosecond spectroscopy},
	issn = {0301-0104},
	doi = {10.1016/j.chemphys.2012.01.017},
	url = {https://www.sciencedirect.com/science/article/pii/S0301010412000298},
	author = {Dahlstr\"om, J. M. and Gu\'enot, D. and Kl\"under, K. and Gisselbrecht, M. and Mauritsson, J. and L'Huillier, A. and Maquet, A. and Ta\"ieb, R.}
}

@article{boll2022PRA,
	title = {Analytical model for attosecond time delays and Fano's propensity rules in the continuum},
	author = {Boll, D. I. R. and Martini, L. and Foj\'on, O. A.},
	journal = {Phys. Rev. A},
	volume = {106},
	issue = {2},
	pages = {023116},
	year = {2022},
	month = {Aug},
	publisher = {American Physical Society},
	doi = {10.1103/PhysRevA.106.023116},
	url = {https://link.aps.org/doi/10.1103/PhysRevA.106.023116}
}

@article{boll2023PRA,
	title = {Two-color polarization control of angularly resolved attosecond time delays},
	author = {Boll, D. I. R. and Martini, L. and Palacios, A. and Foj\'on, O. A.},
	journal = {Phys. Rev. A},
	volume = {107},
	issue = {4},
	pages = {043113},
	year = {2023},
	month = {Apr},
	publisher = {American Physical Society},
	doi = {10.1103/PhysRevA.107.043113},
	url = {https://link.aps.org/doi/10.1103/PhysRevA.107.043113}
}

@article{natalense1999JCP,
	author  = {Natalense, Alexandra P. P. and Lucchese, Robert R.},
	title   = {Cross section and asymmetry parameter calculation for sulfur 1s photoionization of SF$_6$},
	journal = {The Journal of Chemical Physics},
	volume  = {111},
	number  = {12},
	pages   = {5344--5348},
	year    = {1999},
	month   = {Sep},
	doi     = {10.1063/1.479794},
	url     = {https://doi.org/10.1063/1.479794}
}

@article{kendall1992JCP,
	author  = {Kendall, Rick A. and Dunning, Thom H., Jr. and Harrison, Robert J.},
	title   = {Electron affinities of the first‐row atoms revisited. Systematic basis sets and wave functions},
	journal = {The Journal of Chemical Physics},
	volume  = {96},
	number  = {9},
	pages   = {6796--6806},
	year    = {1992},
	month   = {May},
	doi     = {10.1063/1.462569}
}

@article{han2025Nature,
	author  = {Han, Meng and Ji, Jia-Bao and Blech, Alexander and Goetz, R. Esteban and Allison, Corbin and Greenman, Loren and Koch, Christiane P. and W{\"o}rner, Hans Jakob},
	title   = {Attosecond control and measurement of chiral photoionization dynamics},
	journal = {Nature},
	volume  = {645},
	pages   = {95--100},
	year    = {2025},
	doi     = {10.1038/s41586-025-09455-4},
	url     = {https://www.nature.com/articles/s41586-025-09455-4}
}

@article{MendesPCCP2026,
	author = {Mendes, Mónica and Ameixa, João and Rodrigues, Rodrigo and Sequeira, Diogo and Jones, Nykola C. and Hoffmann, Søren V. and Barbosa, Alessandra Souza and da Silva, Filipe Ferreira},
	title = {Probing the photoabsorption features and electronic excited states of propylene oxide: an experimental and theoretical study},
	journal = {Physical Chemistry Chemical Physics},
	volume = {28},
	number = {12},
	pages = {7645-7654},
	year = {2026},
	month = {03},
	issn = {1463-9076},
	doi = {10.1039/d6cp00031b}
}

@article{gallegos1961JACS,
	author  = {Gallegos, E. J. and Kiser, R. W.},
	title   = {Electron Impact Spectroscopy of Ethylene Oxide and Propylene Oxide},
	journal = {J. Am. Chem. Soc.},
	volume  = {83},
	number  = {4},
	pages   = {773--777},
	year    = {1961},
	doi     = {10.1021/ja01465a004}
}

@article{WATANABE1962JQSRT,
	author = {K. Watanabe and Toshio Nakayama and Joseph Mottl},
	title = {Ionization potentials of some molecules},
	journal = {Journal of Quantitative Spectroscopy and Radiative Transfer},
	volume = {2},
	number = {4},
	pages = {369-382},
	year = {1962},
	issn = {0022-4073},
	doi = {https://doi.org/10.1016/0022-4073(62)90023-7}
	
}

@article{McAlduff1977CJC,
	author = {McAlduff, E. J. and Houk, K. N.},
	title = {Photoelectron spectra of substituted oxiranes and thiiranes. Substituent effects on ionization potentials involving σ orbitals},
	journal = {Canadian Journal of Chemistry},
	volume = {55},
	number = {2},
	pages = {318-332},
	year = {1977},
	doi = {10.1139/v77-048}
}

@book{kimura1981JSSP,
	author    = {Kimura, K. and Katsumata, S. and Achiba, Y. and Yamazaki, T. and Iwata, S.},
	title     = {Handbook of {HeI} Photoelectron Spectra of Fundamental Organic
	Molecules: Ionization Energies, Ab Initio Assignments, and
	Valence Electronic Structure for 200 Molecules},
	publisher = {Japan Scientific Societies Press},
	address   = {Tokyo},
	year      = {1981},
	isbn      = {0470272007}
}

@article{liu1999JPCA,
	author  = {Liu, F. and Sheng, L. and Qi, F. and Gao, H. and Li, C. and
	Zhang, Y. and Yu, S. and Lau, K.-C. and Li, W.-K.},
	title   = {A Vacuum Ultraviolet Photoionization Mass Spectrometric Study of
	Propylene Oxide in the Photon Energy Region of 10--40 eV},
	journal = {J. Phys. Chem. A},
	volume  = {103},
	number  = {41},
	pages   = {8179--8186},
	year    = {1999},
	doi     = {10.1021/jp990635o}
}

@article{rodrigues2024JPCA,
	author  = {Rodrigues, R. and Bou Debes, D. and Mendes, M. and Guerra, P. and
	Mestre, G. and Eden, S. and Cornetta, L. M. and Ing{\'o}lfsson, O. and
	Ferreira da Silva, F.},
	title   = {Experimental and Theoretical Study on Electron Ionization and
	Fragmentation of Propylene Oxide---the First Chiral Molecule
	Detected in the Interstellar Medium},
	journal = {J. Phys. Chem. A},
	volume  = {128},
	number  = {24},
	pages   = {4795--4805},
	year    = {2024},
	doi     = {10.1021/acs.jpca.4c02116}
}
\bibliographystyle{apsrev4-2}
	
\end{document}